\documentclass[noeprint,twocolumn,aps,prb,showpacs,groupedaddress,superscriptaddress,nofootinbib,raggedbottom]{revtex4-2} 
\usepackage{graphicx}  
\usepackage[dvipsnames]{xcolor} 
\usepackage{color}     
\usepackage{amsmath}   
\usepackage{amsfonts}  
\usepackage{float}     
\usepackage[pdftex,linkcolor=blue,citecolor=blue,urlcolor=blue,colorlinks]{hyperref} 
\usepackage{outlines} 
\usepackage{lipsum}
\usepackage{bm}    
\usepackage{mathrsfs}
\usepackage{bigstrut}

\newcommand\varpm{\mathbin{\vcenter{\hbox{%
				\oalign{\hfil$\scriptstyle+$\hfil\cr
					\noalign{\kern-.3ex}
					$\scriptscriptstyle({-})$\cr}%
}}}}

\graphicspath{{Figure/}}        
\usepackage{etoolbox}

\begin{document}

\title{Flat-band induced local Hilbert space fragmentation}

\author{Eul\`{a}lia Nicolau}
\affiliation{%
	Departament de F\'{i}sica, Universitat Aut\`{o}noma de Barcelona, E-08193 Bellaterra, Spain.}%

\author{Anselmo M. Marques}%
\affiliation{Department of Physics and i3N, University of Aveiro, 3810-193 Aveiro, Portugal.}%

\author{Ricardo G. Dias}%
\affiliation{Department of Physics and i3N, University of Aveiro, 3810-193 Aveiro, Portugal.}%

\author{Ver\`{o}nica Ahufinger}%
\affiliation{%
	Departament de F\'{i}sica, Universitat Aut\`{o}noma de Barcelona, E-08193 Bellaterra, Spain.}%

\begin{abstract} 
We demonstrate that a complete class of flat-band lattices with underlying commutative local symmetries exhibit a locally fragmented Hilbert space.
The equitable partition theorem ensures distinct parities for the compact localized states (CLSs) present in this class of flat-band lattices and the extended eigenstates of the system. In the presence of on-site bosonic interactions, such models exhibit a conserved quantity, the parity of the number of particles located in all the CLSs in a unit cell. As a consequence, the Hilbert space presents local fragmentation, which is only revealed upon rotating the basis of the Hamiltonian that decouples the CLSs at the single-particle level. We find that the fragmentation is strong and also robust to the addition of long-range interactions. As an example, we numerically analyze the fragmentation of the one-dimensional Pyrochlore chain, which exhibits both nonintegrable sectors, effective single-particle sectors, and frozen states. We also show that the entanglement entropies form a nested-dome structure typical of these fragmented systems and that thermalization is restricted to each sub-sector.
\end{abstract}
\maketitle

\section{Introduction}\label{SecIntroduction}
Compact localized states (CLSs) are eigenstates of a Hamiltonian that have nonzero amplitudes on (typically few) close-by sites and whose amplitude strictly vanishes on the rest \cite{Flach2014a,Maimati2017}. CLSs arise due to geometrical frustration through the interplay between the geometry and the tunneling amplitudes of the model. If the system is periodic, CLSs lead to a macroscopic number of degenerate eigenstates that constitute a flat band. Flat bands have an energy independent of the quasimomentum, such that transport is strongly suppressed. They can generally be related to the presence of CLSs, as these can be constructed as a superposition of degenerate Bloch states \cite{Rhim2019}. 

There are multiple methods of construction to generate flat bands, such as the use of Fano lattices \cite{Flach2014a}, origami rules \cite{Dias2015}, fractals \cite{Nandy2015,Pal2018}, bipartite graphs \cite{Ramachandran2017}, and others \cite{MoralesInostroza2016,Mielke1991,Morfonios2021}. More general procedures also exist, such as solving inverse eigenvalue problems \cite{Maimati2017,Maimati2021a,Maimati2021b,Maimati2021c} or performing band engineering \cite{Xu2015}. Although there is no framework capable of generating all systems known to exhibit flat bands, many CLSs arise as a result of local reflection symmetries in the Hamiltonian. For this class of CLSs, a general formalism has been proposed using the equitable partition theorem (EPT)  from graph theory \cite{Trudeau1994}, and its generalization to complex matrices \cite{Barrett2017,Francis2017,Thune2016}. This theorem allows one to link the presence of commutative local symmetries to the presence of CLSs \cite{Rontgen2018}.

Local Hilbert space fragmentation has been recently shown to arise in a family of diamond necklace lattices with on-site bosonic interactions, which possess local reflection symmetries and single-particle CLSs \cite{Nicolau2022a}. The interplay between CLSs and interactions leads to the appearance of a conserved quantity that fragments the Hilbert space into exponentially many disconnected sectors: the parity of the number of particles in each CLS. In locally fragmented systems, the conserved quantities that shatter the Hilbert space are strictly local \cite{Buca2022,Mukherjee2021,Danieli2020,Hahn2021,Chertkov2021}, in analogy with the conserved local quantities in disorder-free localization \cite{Smith2017} and in lattice gauge theories \cite{Smith2018,Russomanno2020b,Halimeh2022,Borla2020}. In contrast, standard fragmentation is due to the presence of the recently coined crypto-local conserved quantities, those that cannot be expressed as sums of local operators \cite{Buca2023}. 

A natural question arises: Is there a general local fragmentation mechanism that arises in flat-band lattices? In this paper, we answer this question affirmatively for arbitrary flat-band lattices possessing commutative local symmetries associated with local reflection symmetries, and thus obeying the EPT theorem.

The rest of the article is organized as follows. In Sec.~\ref{SecFlatbands}, we define a class of flat-band systems with commutative local symmetries, and in Sec.~\ref{SecDemonstration}, we demonstrate that they exhibit strong local Hilbert space fragmentation in the presence of on-site bosonic interactions. In Sec.~\ref{SecNN}, we discuss the effect of long-range interactions on the conserved quantities. Finally, we provide a numerical example in Sec.~\ref{SecNumerical} and discuss the conclusions in Sec.~\ref{SecConclusion}.

\begin{figure*}
	\includegraphics[width=1.645\columnwidth]{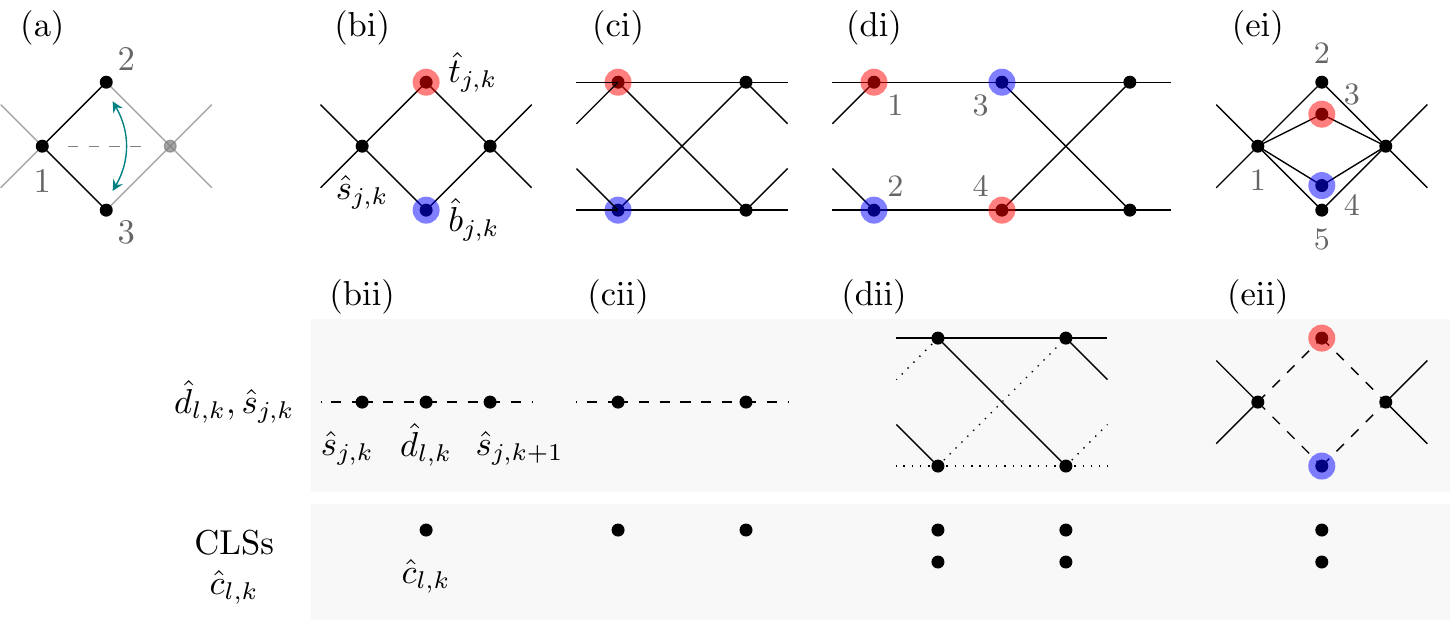}
	\includegraphics[width=0.338\columnwidth]{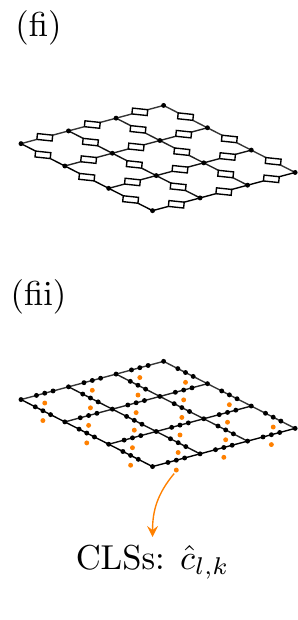}
	\caption{(a) Adjacency graph exhibiting a commutative local symmetry $\mathcal{S}=(1)(2,3)$ which corresponds to a local reflection of the sites $2$ and $3$, where the sites of a unit cell are shown in black. (i) Examples of flat-band lattices with underlying basic commutative local symmetries of order two and (ii) their rotated models composed of dispersive states and spinal sites (top row) and CLSs decoupled at the single-particle level (bottom row). (b) Diamond chain, (c) Creutz ladder, (d) 1D Pyrochlore chain, (e) double diamond chain (f) 2D diamond necklace lattice. Examples of CLSs for the lattices (b)-(e) are given in color, with the radius representing the amplitude and the color representing the phase, where red is a $\pi$ phase and blue is a phase zero. We also indicate the annihilation operators associated with the spinal, top, bottom, dispersive, and CLS sites for a given unit cell $k$.}
	\label{FigScheme}
\end{figure*}

\section{Flat-band lattices from commutative local symmetries}\label{SecFlatbands}
Let us consider the real symmetric matrix associated with a time-reversal invariant and Hermitian Hamiltonian $\hat{\mathcal{H}}$. One might interpret this matrix as an adjacency matrix representing an undirected weighted graph that might contain loops. The vertices of such a graph represent the basis states of the Hamiltonian, while the edges are the adjacency relations between the vertices, i.e. the non-zero matrix elements of $\hat{\mathcal{H}}$. Graphs might present automorphisms, permutations of vertices such that the adjacency relations of the associated matrix are left invariant. Let us take for example a single unit cell of the diamond chain, shown in black in Fig.~\ref{FigScheme}(a), where we have numbered the vertices as 1, 2, and 3. The permutation of the vertices (or sites) $2$ and $3$ leaves the graph invariant. This automorphism can be represented as a permutation that in cyclic notation reads as follows  
\begin{equation}
	\mathcal{S}=(1)(2,3),
\end{equation}
where each parenthesis indicates an \textit{orbit} whose size is the number of elements it contains. An orbit of size $1$ is called a trivial orbit, as it leaves its element unchanged. A permutation $\mathcal{S}$ can also be represented in matrix form $\Pi_\mathcal{S}$, which in the above example would read, in the ordered $\{|1\rangle,|2\rangle,|3\rangle\}$ basis,
\begin{equation}
\Pi_\mathcal{S}=\begin{pmatrix}
	1&0&0\\
	0&0&1\\
	0&1&0
\end{pmatrix}.
\end{equation}
One might consider an enlarged Hamiltonian by taking the black sites that form a diamond in Fig.~\ref{FigScheme}(a) as a unit cell and constructing a lattice  from it [Fig.~\ref{FigScheme}(a) including the gray sites and couplings]. We will adopt the definition of a \textit{commutative local symmetry} of a Hamiltonian $\hat{\mathcal{H}}$ proposed in \cite{Rontgen2018} as a permutation symmetry $\mathcal{S}$ that fulfills
\begin{equation}
	\hat{\mathcal{H}}_{p,q}=\hat{\mathcal{H}}_{\mathcal{S}(p),\mathcal{S}(q)} \, \forall  p,q \quad\Longleftrightarrow\quad [\hat{\mathcal{H}},\Pi_\mathcal{S}]=0.
\end{equation}
A commutative local symmetry is \textit{basic} and of order $o$ if all the non-trivial orbits of $\mathcal{S}$ have the same size $o$.  For basic commutative local symmetries of order two, the permutation matrix fulfills $\Pi_\mathcal{S}^2=\mathbb{I}$. Then, if $\mathcal{S}$ is a commutative local symmetry, the eigenstates of $\hat{\mathcal{H}}$ are also eigenstates of $\Pi_\mathcal{S}$ and have a well-defined parity $\pm1$ with respect to this symmetry.

The equitable partition theorem (EPT) provides a symmetry-induced decomposition of a matrix associated with a graph exhibiting an automorphism into a direct sum of smaller matrices that determine its spectrum and eigenstates \cite{Trudeau1994}. It was originally stated for unweighted graphs, which can be represented by unweighted adjacency matrices, but was later generalized to complex square matrices, thus representing generic Hamiltonians \cite{Barrett2017,Thune2016}. There are two consequences of the EPT that are of interest here. The EPT states that a system with a commutative local symmetry has two classes of eigenstates: eigenstates that are symmetric under the action of $\mathcal{S}$ and eigenstates that are not symmetric and that have support only on the permuted sites \cite{Barrett2017,Francis2017}. Therefore, the EPT ensures the presence of one or more CLSs (depending on the geometry of the Hamiltonian). If the commutative local symmetry $\mathcal{S}$ is basic and of order two, these CLSs will have a well-defined negative parity while all the other states will have a positive parity. Such a permutation can be interpreted visually as a local reflection symmetry with respect to an axis in the adjacency graph of the Hamiltonian [as shown in Fig.~\ref{FigScheme}(a) for the diamond chain]. This interpretation depends on the depiction of the Hamiltonian as a graph but provides an intuitive picture of the destructive interference mechanism that generates the CLSs.

In this work, we will consider lattice models exhibiting CLSs that stem from commutative local symmetries of order two in each unit cell. We represent an arbitrary lattice model with $n$ sites per unit cell as a set of pairs of sites, that we call top and bottom and form a rung. These make explicit the permutation symmetries of the Hamiltonian as a local $y$-reflection symmetry in the Hamiltonian graph. Additionally, each unit cell may present spinal sites that remain invariant under the reflection. Our lattice will have $n=n_1+n_2$ sites per unit cell, where $n_1$ is the number of spinal sites, with associated annihilation operators $\hat{s}_{j,k}$ ($j=1,...,n_1$), and $n_2$ is the number of top and bottom sites, with annihilation operators $\hat{t}_{j,k}$  and $\hat{b}_{j,k}$ ($j=1,...,n_2/2$), respectively. The EPT theorem ensures that such a system presents $n_2/2$ negative-parity CLSs that have support only on the top and bottom sites. Some examples of lattices containing this class of CLSs and an example of their CLSs are given in the top row of Fig.~\ref{FigScheme}: (bi) diamond chain, (ci) Creutz ladder, (di) one-dimensional (1D) Pyrochlore chain, (ei) double diamond chain (fi) two-dimensional (2D) diamond necklace lattice. Note that while some of these lattices [(ai) and (bi)] present one CLS per unit cell, the 1D Pyrochlore chain presents two and the double diamond chain presents three. Also, the double diamond chain can be recast into a top-bottom configuration by reordering the sites. Other examples not depicted in Fig.~\ref{FigScheme} include the square root versions of the diamond chain \cite{Marques2021} or the 2D Creutz ladder \cite{McClarty2020}.

\section{Hilbert space fragmentation}\label{SecDemonstration}
One can also classify CLSs phenomenologically in terms of the number of unit cells that they occupy. In many lattice models, CLSs extending to more than a single unit cell appear as a consequence of inserting a finite magnetic flux per plaquette \cite{Lopes2011,Pelegri2019,Pelegri2019d,Mukherjee2020,Nicolau2022}. In these models, adjacent CLSs of a flat band have spatial overlap, which can be used to generate interaction-driven dynamics \cite{Tovmasyan2013} and topological effects \cite{Pelegri2020,Kuno2020a} in many-body systems. CLSs that occupy a single unit cell can form an orthogonal basis that allows one to detangle each CLSs at the single-particle level \cite{Flach2014a}. For CLSs with underlying commutative local symmetries, this is ensured by the EPT theorem, which restricts the support of the CLSs to the permuted sites. Thus, there is no overlap with the CLSs in adjacent unit cells and they can form an orthogonal basis. Negative parity CLSs, those with an associated basic commutative local  symmetry $\mathcal{S}$ of order two,  constitute a new basis that is completed by their symmetric counterparts. The corresponding annihilation operators are
\begin{equation}\label{EqCLSandDispersive}
	\hat{c}_{l,k}=\sum_{j=1}^{n_2/2} \alpha^l_{j,k}\left(\hat{t}_{j,k}-\hat{b}_{j,k}\right)\!, \,\, 
	\hat{d}_{l,k}=\sum_{j=1}^{n_2/2} \alpha_{j,k}^l\left(\hat{t}_{j,k}+\hat{b}_{j,k}\right)\!,
\end{equation}
where $k$ labels the unit cell, $j$ labels the rung, and the coefficients $\alpha^l_{j,k}$ determine the amplitudes at each rung for each CLS and dispersive state $l=1,...,n_2/2$ in unit cell $k$. In this basis, the CLSs become decoupled at the single-particle level while the symmetric states, which we refer to as dispersive states, remain coupled and compose a dispersive chain supporting extended states. In our notation, the dispersive chain can also include spinal sites, which remain invariant under this rotation. Figures \ref{FigScheme}(ii) represent the rotated models of the (bii) diamond chain, (cii) Creutz ladder, (dii) 1D Pyrochlore chain, (eii) double diamond chain, and (fii) 2D diamond necklace lattice. Each model presents one or more decoupled CLSs for each unit cell. 
 
The many-body Hamiltonian reads $\hat{\mathcal{H}}=\mathcal{J}\hat{\mathcal{H}}^{0}+U \hat{\mathcal{H}}^{int}$, where $\mathcal{J}$ and $U$ indicate the magnitudes of the independent-particle Hamiltonian and the interaction Hamiltonian, respectively. The term $\hat{\mathcal{H}}^0$ can be written as a sum of local operators in each unit cell $\hat{\mathcal{H}}^0=\sum_k \hat{f}_k$, where the operators $\hat{f}_k$ include particle-conserving products of the operators $\hat{\nu}_{j,k}$, with $\nu=t,b,s$. We consider the addition of on-site bosonic interactions of the form $\hat{n}_{\nu_{j,k}}(\hat{n}_{\nu_{j,k}}-1)$, where $\hat{n}_{\nu_{j,k}}=\hat{\nu}_{j,k}^{\dagger}\hat{\nu}_{j,k}$, are the number operators at each site. The interaction Hamiltonian at the top and bottom sites can be written as
\begin{equation}
		\hat{\mathcal{H}}^{int}_{t,b}=\sum_k \sum_{j=1}^{n_2/2} \left[ \hat{t}_{j,k}^{\dagger}\hat{t}_{j,k}^{\dagger}\hat{t}_{j,k}\hat{t}_{j,k}+\hat{b}_{j,k}^{\dagger}\hat{b}_{j,k}^{\dagger}\hat{b}_{j,k}\hat{b}_{j,k}\right].
\end{equation}

It has been recently shown that the parity of the number of particles in each CLS commutes with the Hamiltonian of a diamond chain \cite{Nicolau2022a}. However, the associated conservation quantity remains hidden in the product state basis. We denote the rotated basis using its associated collection of annihilation operators, $\{\hat{c}_{l,k},\hat{d}_{l,k}\}$, and the product state basis as $\{\hat{t}_{j,k},\hat{b}_{j,k}\}$. We will show that a similar mechanism occurs for arbitrary flat-band lattices with basic commutative local symmetries of order two, which may have more than one CLS per unit cell. In this general case, we postulate that the conserved quantity is the parity of the number of particles \textit{in all CLSs of a unit cell}. Thus, the operator 
\begin{equation}
	\hat{\mathcal{P}}_k=e^{i\pi\sum_{l=1}^{n_2/2} \hat{n}_{c_{l,k}}}	
\end{equation}
commutes with the Hamiltonian, $[\hat{\mathcal{P}}_k,\hat{\mathcal{H}}^\prime]=0$, where $\hat{\mathcal{H}}^\prime$ is the total Hamiltonian in the rotated basis, $\hat{n}_{c_{l,k}}=\hat{c}_{l,k}^{\dagger}\hat{c}_{l,k}$, and $n_2/2$ is the number of CLSs in each unit cell. This conservation law leads to the fragmentation of the Hilbert space in the rotated or entangled basis, while it remains hidden in the product-state basis. Therefore, this is an instance of quantum Hilbert space fragmentation, also shown to appear in Temperley-Lieb chains \cite{Moudgalya2021b}, the quantum East model \cite{Brighi2022b}, and spin $1/2$ chains with hard rod deformations \cite{Borsi2023}. In contrast to quantum fragmentation, most examples of Hilbert space fragmentation are classical, leading to a fragmentation structure that is reproducible in classical Markov generators \cite{Moudgalya2021b,Moudgalya2021a}.

Given that the CLSs only have support on the top and bottom sites, the spinal part of the interaction Hamiltonian trivially commutes with $\hat{\mathcal{P}}_k$. The rotated interaction Hamiltonian on the top and bottom sites can be obtained by expressing the top and bottom operators in terms of CLSs and dispersive operators [Eq.~(\ref{EqCLSandDispersive})]. Due to the periodicity of the lattice, it is enough to consider a single unit cell $k$. Thus, we omit the unit cell index $k$ in the demonstration, for ease of reading. The reduced interaction Hamiltonian in the rotated basis reads
\begin{equation}\label{EqInteractionRotated}
	\begin{aligned}
	\hat{\mathcal{H}}^{int\prime}_{t,b}&=\sum_{\hat{\rho},\hat{\sigma},\hat{\tau},\hat{\upsilon}}\sum_{\{l_i\}=1}^{n_2/2}\left(\Theta^{l_1,l_2,l_3,l_4}_{\hat{\rho},\hat{\sigma},\hat{\tau},\hat{\upsilon}}\right)\hat{\rho}_{l_1}^{\dagger}\hat{\sigma}_{l_2}^{\dagger}\hat{\tau}_{l_3}\hat{\upsilon}_{l_4},
\end{aligned}
\end{equation}
where $\{\hat{\rho},\hat{\sigma},\hat{\tau},\hat{\upsilon}\}=\{\hat{c},\hat{d}\}$ are the annihilation operators of either a CLS or a dispersive state, respectively, $\{l_i\}=1,...,n_2/2$ are the CLS and dispersive state indices (with $i=1,2,3,4)$, and $\Theta^{l_1,l_2,l_3,l_4}_{\hat{\rho},\hat{\sigma},\hat{\tau},\hat{\upsilon}}$ are the coefficients of each term. Only those terms with an odd number of $\hat{c}_{l}^{(\dagger)}$ and of $\hat{d}_{l}^{(\dagger)}$ operators do not commute with $\hat{\mathcal{P}}_k$. Those terms, e.g. $\hat{c}_{l_1}^{\dagger}\hat{d}_{l_2}^{\dagger}\hat{d}_{l_3}\hat{d}_{l_4}$, exchange one particle between the CLSs and the dispersive states, thus violating parity. Then, it is enough to prove that the coefficient $\Theta^{l_1,l_2,l_3,l_4}_{\hat{\rho},\hat{\sigma},\hat{\tau},\hat{\upsilon}}$ vanishes for terms of this form.

The rotation matrix from the product-state basis $\{\hat{t}_j,\hat{b}_j\}$  to the rotated basis $\{\hat{c}_l,\hat{d}_l\}$ in each unit cell can be written as a Kronecker product $\mathcal{K}=\mathcal{L}\otimes \mathcal{M}$, with 
\begin{equation}
	\mathcal{L}=\begin{pmatrix}
		\alpha_1^1  & \alpha_2^1 &\\
		\alpha_1^2  & \alpha_2^2 & \dots\\
		&\vdots&\ddots
	\end{pmatrix} \qquad \mbox{and} \qquad \mathcal{M}=\begin{pmatrix}1  & 1 \\-1  & 1 \\ \end{pmatrix}.
\end{equation}
where we have ordered the basis as $\{\hat{b}_1,\hat{t}_1,\hat{b}_2,\hat{t}_2,...\}$ and $\{\hat{d}_1,\hat{c}_1,\hat{d}_2,\hat{c}_2,...\}$, and made use of Eq.~(\ref{EqCLSandDispersive}). The inverse of matrix $\mathcal{K}$ is $\mathcal{K}^{-1}=(\mathcal{L}\otimes \mathcal{M})^{-1}=\mathcal{L}^{-1}\otimes \mathcal{M}^{-1}$, and indicates the expressions of $\{\hat{t}_j,\hat{b}_j\}$ in terms of $\{\hat{c}_l,\hat{d}_l\}$. By writing $\mathcal{L}^{-1}$ as an arbitrary matrix, $\mathcal{K}^{-1}$ takes the following general form 
\begin{equation}
	\mathcal{K}^{-1}=\begin{pmatrix}
		\beta_1^1   & -\beta_1^1   & \beta_2^1   & -\beta_2^1  &\\
		\beta_1^1   & \beta_1^1   & \beta_2^1   & \beta_2^1 &\\
		\beta_1^2   & -\beta_1^2   & \beta_2^2   & -\beta_2^2  &\\
		\beta_1^2   & \beta_1^2   & \beta_2^2   & \beta_2^2  &\dots\\
		&&&\vdots&\ddots
	\end{pmatrix}.
\end{equation}
Thus, one can express the annihilation operators $\hat{t}_j$ and $\hat{b}_j$ in terms of $\hat{c}_l$ and $\hat{d}_l$
\begin{equation}
	\hat{t}_{j,k}=\sum_{l=1}^{n_2/2} \beta^j_{l,k}\left(\hat{d}_{l,k}+\hat{c}_{l,k}\right)\!, \,\,
	\hat{b}_{j,k}=\sum_{l=1}^{n_2/2} \beta_{l,k}^j\left(\hat{d}_{l,k}-\hat{c}_{l,k}\right)\!.
\end{equation}
Then, the coefficient $\Theta^{l_1,l_2,l_3,l_4}_{\hat{\rho},\hat{\sigma},\hat{\tau},\hat{\upsilon}}$ for the terms containing an odd number of CLS and dispersive operators takes two forms: ($i$) for the terms with one CLS operator and three dispersive state operators and ($ii$) for the terms with three CLS operators and one dispersive state operator. These two terms are
\begin{equation}
	\begin{aligned}
		(i)&\quad \sum_{j=1}^{n_2/2}\left[\beta_r^j\beta_m^j\beta_p^j\beta_o^j+(-\beta_r^j)\beta_m^j\beta_p^j\beta_o^j\right]=0,\\
		(ii) &\quad\sum_{j=1}^{n_2/2}\left[\beta_r^j\beta_m^j\beta_p^j\beta_o^j+\beta_r^j(-\beta_m^j)(-\beta_p^j)(-\beta_o^j)\right]=0.
	\end{aligned}
\end{equation}
For both cases, it vanishes at each rung $j$. As all the other terms commute with $\mathcal{\hat{P}}_k$, the parity of the number of particles in all the CLSs in a unit cell is conserved. The only terms in the rotated Hamiltonian that produce a particle exchange between the CLSs and the dispersive states are of the form $\hat{d}_{l_1,k}^{\dagger}\hat{d}_{l_2,k}^{\dagger}\hat{c}_{l_3,k}\hat{c}_{l_4,k}$ and $\hat{c}_{l_1,k}^{\dagger}\hat{c}_{l_2,k}^{\dagger}\hat{d}_{l_3,k}\hat{d}_{l_4,k}$. These denote a two-particle tunneling between CLSs and dispersive states that preserves $\hat{\mathcal{P}}_k$. This mechanism is a direct consequence of the commutative local symmetry of these lattices: the well-defined parities of the eigenstates determine the structure of the basis states (\ref{EqCLSandDispersive}) which in turn determines the form of $\mathcal{M}^{-1}$. Note that this result is not restricted to one dimension, as the underlying permutations can exchange sites in any axis [see Fig.~\ref{FigScheme}(f)].

As the operator $\hat{\mathcal{P}}_k$ in each unit cell $k$ commutes with the rotated Hamiltonian $\hat{\mathcal{H}}^\prime$, one can also define the total CLS number parity as $\hat{\mathcal{P}}=\sum_k\hat{\mathcal{P}}_k$, which also commutes with $\hat{\mathcal{H}}^\prime$. The rotated Hamiltonian is then composed of a series of sectors defined by the eigenvalues of $\hat{\mathcal{P}}$ and within those, one or more sub-sectors determined by the eigenvalues of $\hat{\mathcal{P}}_k$. The eigenvalues of $\hat{\mathcal{P}}$ and $\hat{\mathcal{P}}_k$ are given in Table \ref{Table} as well as the number of sectors and sub-sectors in terms of the number of particles $N$ and the number of unit cells $N_c$. The number of sub-sectors grows exponentially with system size, signaling Hilbert space fragmentation \cite{Moudgalya2021b}. In particular, this mechanism produces \textit{local} Hilbert space fragmentation, as the shattering of the Hilbert space stems from a strictly local conservation law, $[\hat{\mathcal{P}}_k,\hat{\mathcal{H}}^\prime]=0$ \cite{Buca2022}.

The degree of fragmentation can be measured by calculating the ratio of the dimension of the largest sector of the Hilbert space to the total dimension of the space \cite{Sala2020}. For our class of models, the dimension of the largest sub-sector is 
\begin{equation}\begin{aligned}
		\mathscr{D}_{max}\hspace{-0.7mm}=&\sum_{\varrho=0}^{\left\lfloor \frac{N}{2}\right\rfloor}\Bigg[\binom{(n_1+\frac{n_2}{2})N_c+N-2\varrho-1}{N-2\varrho}\hspace{-1.2mm}\\
		&\times\sum_{(\delta_k,...,\delta_{N_c})\in\mathcal{Q}}\,\,\prod_{k=1}^{N_c}\binom{\frac{n_2}{2}+2\delta_k-1}{2\delta_k}\hspace{-0.5mm}\Bigg],
	\end{aligned}
\end{equation}
where the indices $\varrho$ and $\delta_k$ count the number of pairs of particles that populate the CLSs, in total and for a unit cell $k$, respectively, $N$ is the number of particles, and the set $\mathcal{Q}$ fulfills $\mathcal{Q}(\varrho,N_c)=\{(\delta_k,...,\delta_{N_c})|\varrho=\delta_1+\delta_2+...+\delta_{N_c}\}$.
 The dimension of the full Hilbert space is
\begin{equation}
	\mathscr{D}=\binom{nN_c+N-1}{N}.
\end{equation}
The ratio $\mathscr{D}_{max}/\mathscr{D}$ tends to zero at the thermodynamic limit, indicating strong Hilbert space fragmentation. By contrast, a limit of one would indicate weak fragmentation, where the largest sector dominates.

Let us consider some examples. For the class of models with only one CLSs per unit cell [Fig.~\ref{FigScheme}(bi) diamond chain and (ci) Creutz ladder], the conserved quantity simplifies to $\hat{\mathcal{P}}_k=e^{i\pi\hat{n}_{c_k}}$. The double diamond chain, Fig.~\ref{FigScheme}(ei), is an unusual example, it presents multiple commutative local symmetries of order two, such as 
\begin{equation}
	\begin{aligned}
	\mathcal{S}_1=(1)(4)(5)(2,3), \quad \mathcal{S}_2=(1)(2)(3)(4,5), \\ \mathcal{S}_3=(1)(2)(5)(3,4), \quad
	\mathcal{S}_4=(1)(2)(4)(3,5).
	\end{aligned}
\end{equation}
Taking for example $\mathcal{S}_1$ and $\mathcal{S}_2$, these are \textit{independent} commutative local symmetries that lead to non-overlapping CLSs and thus to independently conserved quantities. However, after decoupling these CLSs the dispersive  lattice still presents an unresolved  local symmetry [see Fig.~\ref{FigScheme}(eii)]. The third CLS occupies all diamond sites in Fig.~\ref{FigScheme}(ei) and corresponds to the permutation $\mathcal{S}=(1)(2,5)(3,4)$. One can perform a second rotation to decouple this state at the single-particle level. However, the third CLS will not be decoupled from the dispersive chain at the many-body level due to the presence of interaction-induced one-particle tunnelings between the dispersive chain and the CLS. In contrast, the 1D Pyrochlore chain presents the symmetry $\mathcal{S}=(1,2)(3,4)$ [see Fig.~\ref{FigScheme}(di)], which cannot be decomposed into two independent permutations. As a consequence, it presents two overlapping CLSs per unit cell that lead to a single conserved quantity  $\hat{\mathcal{P}}_k=e^{i\pi(\hat{n}_{c_{1,k}}+\hat{n}_{c_{2,k}})}$. Therefore, each independent local reflection symmetry with an underlying basic commutative local symmetry $\mathcal{S}$ of order two leads to a conserved quantity. These require a single rotation to detangle the associated CLSs and thus lead to the conservation of parity and fragmentation. Some lattices, such as the double diamond chain, might present more than one independent local symmetry per unit cell, which leads to a multiplicity of conserved quantities. For example, one might create an enlarged unit cell by uniting  Creutz and 1D Pyrochlore unit cells, which will lead to two independent sets of conserved quantities per unit cell.

\begin{table}[t]
	\begin{center}
		\begin{tabular}{ccc}
			\hline\hline
			& \hspace{-2mm}No. sectors & \hspace{-10.5mm}and sub-sectors\\ 
			Eigenvalue &  $N>N_c$ &  $N\leq N_c$\\ \hline
			\rule{0pt}{12pt}
			$\mathcal{P}_k=\pm1$ & $2^{N_c}$ & \hspace{-1mm}$\sum_{k=0}^{N-1}\binom{N_c}{k}\hspace{-0.7mm}+\hspace{-0.7mm}\binom{N_c}{N}\hspace{-0.5mm}(\frac{n_2}{2})^N$\\[2pt] 
			$\mathcal{P}=-N_c,-N_c+2,...,N_c$ & $N_c+1$ & $N+1$\\\hline\hline 
		\end{tabular}
	\end{center}
	\caption{Eigenvalues of the local, $\hat{\mathcal{P}}_k$, and total, $\hat{\mathcal{P}}$, parities, and number of associated sub-sectors, $\mathcal{P}_k$ and sectors $\mathcal{P}$ for a number of particles $N$ larger, equal, or smaller than the number of unit cells $N_c$.}
	\label{Table}
\end{table}

\section{Long-range interactions}\label{SecNN}
Let us consider how the block-diagonal structure of the Hamiltonian is affected by the presence of long-range interactions. There are mainly three classes of long-range interaction terms for a 1D system. Considering interactions that respect the $y$-reflection symmetry of the system, these read 
\begin{equation}
	\begin{aligned}
		(i)\, &\hat{\mathcal{H}}^{int}_{1}=\sum_{k,k^\prime\hspace{-0.7mm}}\sum_{j,j^\prime}\xi_{k,k^\prime\hspace{-0.7mm}}^{j,j^\prime}\left(\hat{n}_{s_{j,k}}\hat{n}_{t_{j^\prime,k^\prime}}+\hat{n}_{s_{j,k}}\hat{n}_{b_{j^\prime,k^\prime}}\right),\\
		(ii)\, &\hat{\mathcal{H}}^{int}_{2}=\sum_{k}\sum_{j}\vartheta_j\hat{n}_{t_{j,k}}\hat{n}_{b_{j,k}},\\
		(iii)\, &\hat{\mathcal{H}}^{int}_{3}=\sum_{k,k^\prime\hspace{-0.7mm}}\sum_{j,j^\prime}\Omega_{k,k^\prime\hspace{-0.7mm}}^{j,j^\prime}\left(\hat{n}_{t_{j,k}}\hat{n}_{t_{j^\prime,k^\prime}}+\hat{n}_{t_{j,k}}\hat{n}_{b_{j^\prime,k^\prime}}\right. \\  &\left.\hspace{25mm}+\hat{n}_{b_{j,k}}\hat{n}_{t_{j^\prime,k^\prime}}+\hat{n}_{b_{j,k}}\hat{n}_{b_{j^\prime,k^\prime}}\right).
	\end{aligned}
\end{equation}
In term ($iii$), we have assumed that the cross terms, e.g. $\hat{n}_{t_{j,k}}\hat{n}_{b_{j^\prime,k^\prime}}$, have the same strength than the horizontal terms, e.g. $\hat{n}_{t_{j,k}}\hat{n}_{t_{j^\prime,k^\prime}}$. There can also be interactions between spinal sites, which remain invariant under the basis rotation and thus preserve fragmentation. In order to understand the effect of these terms, we can write them in the rotated basis determined by the annihilation operators $\hat{\tilde{c}}_{j,k}=\left(\hat{t}_{j,k}-\hat{b}_{j,k}\right)/\sqrt{2}$, and $\hat{\tilde{d}}_{j,k}=\left(\hat{t}_{j,k}+\hat{b}_{j,k}\right)/\sqrt{2}$. In contrast with the basis considered before [see Eq.~(\ref{EqCLSandDispersive})], the states annihilated by $\hat{\tilde{c}}_{j,k}$ are not eigenstates of the system, i.e., they are not the CLSs, except for the cases of the diamond chain and Creutz ladder, where the CLSs occupy a single rung $j$ [Fig.~\ref{FigScheme}(bi) and (ci)]. Thus, the CLSs do not generally become decoupled through this rotation. For arbitrary lattices, the antisymmetric states represented by $\hat{\tilde{c}}_{j,k}$ are superpositions of the CLSs of unit cell $k$, and thus remain coupled between them within a unit cell at the single-particle level. The symmetric states given by $\hat{\tilde{d}}_{j,k}$ form a dispersive chain that is decoupled from the states given by $\hat{\tilde{c}}_{j,k}$. The rotated interaction Hamiltonians in this basis read
\begin{equation}
	\begin{aligned}
		(i)\, \hat{\mathcal{H}}^{int\prime}_1=&\sum_{k,k^\prime\hspace{-0.7mm}}\sum_{j,j^\prime}\xi_{k,k^\prime\hspace{-0.7mm}}^{j,j^\prime}\left(\hat{n}_{\tilde{d}_{j,k}}\hat{n}_{s_{j^\prime\hspace{-0.7mm},k^\prime\hspace{-0.7mm}}}+\hat{n}_{\tilde{c}_{j,k}}\hat{n}_{s_{j^\prime\hspace{-0.7mm},k^\prime\hspace{-0.7mm}}}\right)\\
		(ii)\,  \hat{\mathcal{H}}^{int\prime}_2=&\sum_{k}\sum_{j}\dfrac{\vartheta_j}{4}\left[\hat{n}_{\tilde{d}_{j,k}}(\hat{n}_{\tilde{d}_{j,k}}-1)+\hat{n}_{\tilde{c}_{j,k}}(\hat{n}_{\tilde{c}_{j,k}}-1)\right.\\&\hspace{15mm}\left.-\hat{\tilde{c}}^\dagger_{j,k}\hat{\tilde{c}}^\dagger_{j,k}\hat{\tilde{d}}_{j,k}\hat{\tilde{d}}_{j,k}-\hat{\tilde{d}}^\dagger_{j,k}\hat{\tilde{d}}^\dagger_{j,k}\hat{\tilde{c}}_{j,k}\hat{\tilde{c}}_{j,k}\right]\\
		(iii)\,  \hat{\mathcal{H}}^{int\prime}_3=&\sum_{k,k^\prime\hspace{-0.7mm}}\sum_{j,j^\prime}\Omega_{k,k^\prime\hspace{-0.7mm}}^{j,j^\prime}\left(\hat{n}_{\tilde{d}_{j,k}}\hat{n}_{\tilde{d}_{j^\prime\hspace{-0.7mm},k^\prime\hspace{-0.7mm}}}+\hat{n}_{\tilde{c}_{j,k}}\hat{n}_{\tilde{c}_{j^\prime\hspace{-0.7mm},k^\prime\hspace{-0.7mm}}}\right.\\[-2mm]
		&\hspace{20mm}\left.+\hat{n}_{\tilde{d}_{j,k}}\hat{n}_{\tilde{c}_{j^\prime\hspace{-0.7mm},k^\prime\hspace{-0.7mm}}}+\hat{n}_{\tilde{c}_{j,k}}\hat{n}_{\tilde{d}_{j^\prime\hspace{-0.7mm},k^\prime\hspace{-0.7mm}}}\right).
	\end{aligned}
\end{equation}
The Hamiltonians ($i$) and ($iii$) are defined for any distance between the first, $k$, and second, $k'$, unit cells, and might involve different pairs of rungs $j,j^\prime$. Thus, they represent not only nearest-neighbour (NN) interactions but arbitrary long-range interactions. These only include density-density interaction terms in the rotated basis. Thus, they conserve the number of particles in all CLSs of a unit cell, as particles are free to move between the states given by $\hat{\tilde{c}}_{j,k}$ of a single unit cell. The Hamiltonian ($ii$) also includes two particle tunnelings between the states annihilated by $\hat{\tilde{c}}_{j,k}$ and $\hat{\tilde{d}}_{j,k}$, such that only the parity of the number of particles in all CLSs of a unit cell is conserved. Therefore, all these terms preserve the fragmentation of the Hilbert space and the parity sectors determined by on-site interactions.  If one considers the case where there are no on-site interactions and only long-range interactions of the form ($i$) and ($iii$), then, the structure of the fragmented Hilbert space changes, as each sub-sector is given by the number of particles in each CLS (not the parity), and the number of sub-sectors proliferates, leading to a stronger fragmentation. As the authors of \cite{Tovmasyan2018} point out, all density-density interactions invariant under the graph automorphism associated with the local symmetry will preserve fragmentation. In particular, they study the Creutz ladder, diamond chain, and dice lattice with flux. Their formalism can be used to analyze the generic class of flat-band lattices with commutative local symmetries studied here, as we show in Appendix \ref{SecIntertwinning}. Similar considerations have been pointed out for the case of all-bands-flat lattices \cite{Danieli2021}.

\begin{figure*}[t]
	\includegraphics[width=2.05\columnwidth]{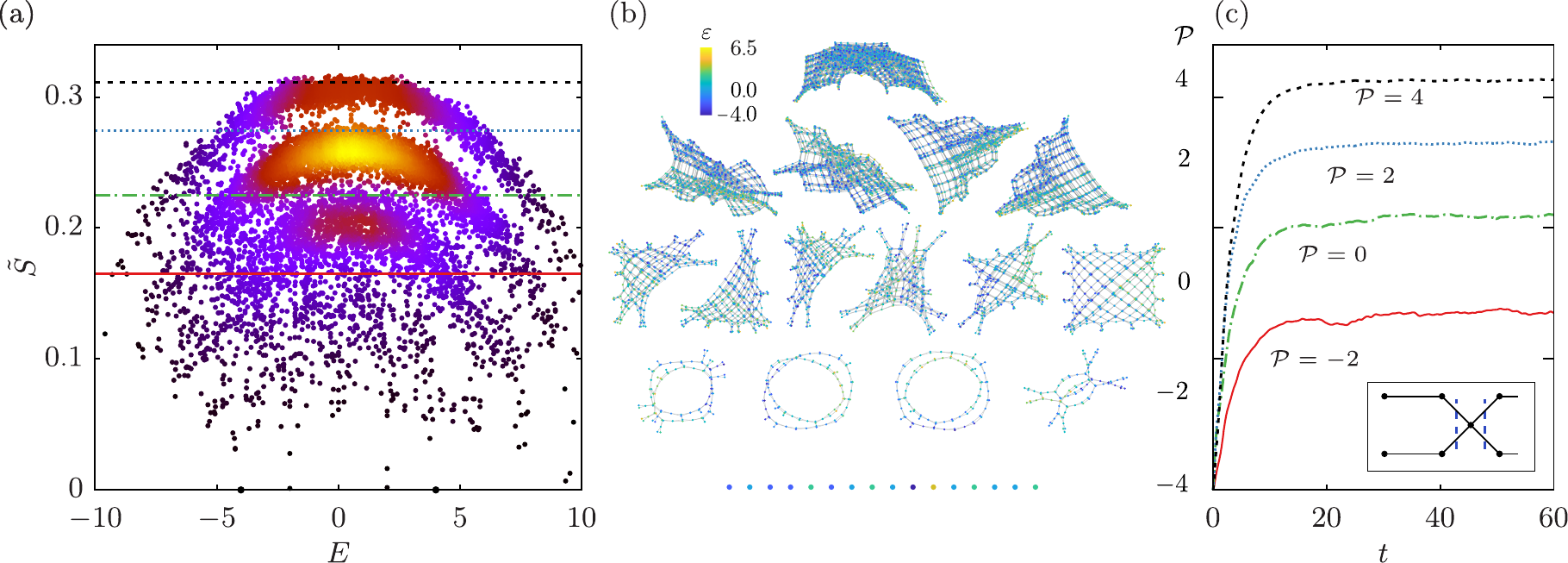}
	\caption{Numerical results  for the 1D Pyrochlore chain with $N=4$ particles in $N_c=4$ unit cells and $\mathcal{J}=U=1$. (a) Half-chain bipartite von Neumann entanglement entropy of each eigenstate as a function of the energy. The horizontal lines are the sector-restricted Page values for each sector and the color of the dots indicates the normalized density of data points, increasing with warming colors. (b) Adjacency graphs corresponding to the dome structures in (a) for each sector $\mathcal{P}$, where the color of the nodes represents the diagonal terms of the rotated Hamiltonian, $\varepsilon$. (c) Average of the entanglement entropy evolution for $20$ random rotated basis states of each sector with eigenvalue $\mathcal{P}$. Inset: beginning of the 1D Pyrochlore chain with one spinal site and bipartitions (dashed blue lines) resulting in a zero entanglement entropy for the CLSs. In (a) and (c), the entropy is normalized to the number of sites in the subsystem, $\tilde{S}=S/\mathcal{N}_L$.}
	\label{FigPyrochlore}
\end{figure*}

\section{Example: Pyrochlore lattice}\label{SecNumerical}
In this section, we numerically study the 1D Pyrochlore chain  [Fig.~\ref{FigScheme}(di)] as an example. Each unit cell contains two CLSs that, together with two dispersive states, form a new basis. The associated annihilation operators are given by
\begin{equation}\label{EqBasisPyrochlore}
\begin{aligned}
	\hat{c}_1=\frac{1}{2}\left(\hat{t}_1+\hat{t}_2-\hat{b}_1-\hat{b}_2\right)\hspace{-1mm}, & \quad	\hat{c}_2=\frac{1}{2}\left(\hat{t}_1-\hat{t}_2-\hat{b}_1+\hat{b}_2\right)\hspace{-1mm},\\
	\hat{d}_1=\frac{1}{2}\left(\hat{t}_1+\hat{t}_2+\hat{b}_1+\hat{b}_2\right)\hspace{-1mm}, & \quad
	\hat{d}_2=\frac{1}{2}\left(\hat{t}_1-\hat{t}_2+\hat{b}_1-\hat{b}_2\right)\hspace{-1mm},
\end{aligned}
\end{equation}
where we have omitted the unit cell index $k$. Note that these CLSs occupy two rungs, in contrast to the states defined by $\hat{\tilde{c}}_{j,k}$ in the previous section. The single-particle energies of the CLSs are $E_{c_1}=1$ and $E_{c_2}=-1$, and thus the states $\hat{\tilde{c}}_{j,k}$ are not eigenstates of the single-particle Hamiltonian. If the CLSs were degenerate, the states given by $\hat{\tilde{c}}_{j,k}$ would also be eigenstates. In this basis, the interaction Hamiltonian contains only terms that commute with $\hat{\mathcal{P}}_k$, thus conserving the parity of the number of particles in the two CLSs, $\hat{\mathcal{P}}_k=e^{i\pi(\hat{n}_{c_{1,k}}+\hat{n}_{c_{2,k}})}$. In a unit cell $k$, it reads  
\begin{equation}\label{EqInteractionPyro}
	\begin{aligned}
		\hat{\mathcal{H}}^{int\prime}=&\dfrac{1}{8}\sum_{l,l'} \left(\hat{d}_{l}^\dagger \hat{d}_{l}^\dagger \hat{c}_{l’}\hat{c}_{l’}+\rm{H.c.}\right)\\
		&+\dfrac{1}{8}\sum_{l, l’}\left( \hat{d}_{l}^\dagger \hat{d}_{l}^\dagger \hat{d}_{l’}\hat{d}_{l’}+\hat{c}_{l}^\dagger \hat{c}_{l}^\dagger \hat{c}_{l’}\hat{c}_{l’}\right)\\
		&+\dfrac{1}{2} \bigg(\hat{n}_{d_{1}}\hat{n}_{d_{2}}+\hat{n}_{c_{1}}\hat{n}_{c_{2}}+\sum_{l, l'}\hat{n}_{d_{l}}\hat{n}_{c_{l’}}\bigg)\\
		&+\dfrac{1}{2}\bigg[\sum_{l,l'} \hat{c}_{l}^\dagger \hat{d}_{l’}^\dagger \hat{c}_{\bar{l}}\hat{d}_{\bar{l}’}\hspace{-0.6mm}+\hspace{-0.6mm}\left(\hat{c}_{1}^\dagger \hat{c}_{2}^\dagger \hat{d}_{1}\hat{d}_{2}+\rm{H.c.}\right)\hspace{-1mm}\bigg].
	\end{aligned}
\end{equation}
where $\bar{l}^{(\prime)}$ indicates the opposite index of $l^{(\prime)}$, i.e., $\bar{l}^{(\prime)}\neq l^{(\prime)}$. The terms include on-site interactions, two-particle tunnelings, and NN interactions in the basis $\{\hat{c}_l,\hat{d}_l\}$. The two-particle tunnelings and NN interactions both include terms between the dispersive states and the CLSs and within these two groups.

Figure~\ref{FigPyrochlore} presents the numerical results for the 1D Pyrochlore chain with $N=4$ particles in $N_c=4$ unit cells, with $\mathcal{J}=U=1$, and open boundary conditions. The lattice starts with the sites hosting the CLSs, presents an integer number of unit cells, and one spinal site [see inset in Fig.~\ref{FigPyrochlore}(c)]. Fig.~\ref{FigPyrochlore}(a) shows the von Neumann half-chain bipartite entanglement entropy of each eigenstate as a function of the energy. We represent the entanglement entropy normalized to the number of sites in the left subsystem, $\tilde{S}=S/\mathcal{N}_L$. The horizontal lines are the sector-restricted Page values \cite{Page1993,Nicolau2022a}, i.e., the average value of $\tilde{S}$ for random states belonging to a particular sector $\mathcal{P}$. The entanglement entropies present a nested-dome structure that can be understood by analyzing the adjacency graph of the many-body Hamiltonian in the rotated basis, shown in Fig.~\ref{FigPyrochlore}(b). The eigenstates that compose each dome correspond to a total sector of the Hamiltonian with eigenvalue $\mathcal{P}$, composed of one or more sub-sectors with eigenvalues $\bm{\mathcal{P}}=(\mathcal{P}_1,\cdots,\mathcal{P}_{N_c})$, where the vector $\bm{\mathcal{P}}$ contains the eigenvalues of $\hat{\mathcal{P}}_k$ at each unit cell $k$. The color of the nodes in Fig.~\ref{FigPyrochlore}(b) represents the diagonal terms of the rotated Hamiltonian, $\varepsilon=\langle f|\hat{\mathcal{H}}^\prime|f\rangle$, where $|f\rangle$ is a basis state, which highlights that the different sub-sectors of a given sector are not degenerate. The entanglement entropies of the CLSs are exactly zero along several bipartitions of the lattice, one of which coincides with the half-chain cut considered in Fig.~\ref{FigPyrochlore}(a). Thus, particles located in a CLS do not contribute to the entanglement entropy of the eigenstates. The two-particle tunneling terms in Eq.~(\ref{EqInteractionPyro}) allow pairs of particles to jump to a CLS. However, those special basis states form a small fraction of the basis states in each sub-sector, and thus the main basis states determine the structure of the entanglement entropies. For the upper dome, most basis states have all particles in a dispersive state, thus obtaining the maximum entanglement entropy. For the lower domes, the presence of basis states with one or more particles in a CLS imposes an upper bound to the entanglement entropy of the corresponding eigenstates. In the sector with $\mathcal{P}=-4$, all particles occupy  CLSs in distinct unit cells. The single particle energies of the CLSs in Eq.~(\ref{EqBasisPyrochlore}) are $E_{c_1}=1$ and $E_{c_2}=-1$, thus, the available energies for the frozen states are $\{-4,-2,0,-2,4\}$, with degeneracies $\{1,4,6,4,1\}$. The two non-degenerate states with energies $\pm4$ have exactly zero entanglement entropy and correspond to the case where all particles populate the same CLS in distinct unit cells. The additional frozen states are degenerate and thus a higher value of $\tilde{S}$ is obtained numerically.

The nested-dome structure of the entanglement entropies is a direct consequence of the fragmentation of the Hilbert space and the low entanglement of the CLSs. Thus, it is generally present in the class of flat-band lattices with commutative local symmetries. However, the distinguishability of the different domes is not guaranteed, as it depends on several factors such as the presence of symmetries, the particle filling, the boundary conditions, and the sparsity of the CLS compared to the dispersive states \cite{Nicolau2022a}. Here, the visibility of the domes is enhanced by introducing one spinal site that makes the CLSs more sparse while also breaking the $x$-reflection symmetry of the model [see inset in Fig.~\ref{FigPyrochlore}(c)]. The nested-dome structure is partly determined by the number of sectors of the system, given in Table \ref{Table}. If there are more particles $N$ than unit cells $N_c$, the number of sectors is always the same, $N_c+1$. However, for each extra particle added with respect to $N=N_c$, a dome is added on top while one sector disappears from the bottom. Additionally, the nested-dome structure slightly shifts to the right as the higher number of particles causes an increase in the interaction energy of the eigenstates since the interactions are assumed to be positive semi-definite. Note that for $N>N_c$ the frozen states are unavailable since there are extra particles that can populate the dispersive chain and are thus free to move. For $N<N_c$, the number of sectors is $N+1$, such that removing particles causes the disappearance of the uppermost domes while the frozen states gain degeneracy due to the multiplicity of the available CLSs.

The Hilbert space presents both integrable and non-integrable sub-sectors. The integrable sub-sectors comprise the frozen states and those in sector $\mathcal{P}=-2$, for which only one particle is free to move in the dispersive chain, thus forming an effective single-particle model. All the other sub-sectors are non-integrable, as they present Wigner-Dyson statistics (numerical evidence is given in Appendix \ref{SecLevelStatistics}). Figure \ref{FigPyrochlore}(c) shows the evolution of the average entanglement entropy for $20$ random initial states belonging to particular sectors. The entanglement entropy grows for all cases while remaining bounded by the sector-restricted Page value indicated in Fig.~\ref{FigPyrochlore}(a). This is a direct consequence of the fragmentation of the Hilbert space, which restricts thermalization within each sub-sector. Such behavior is consistent with the extension of the Eigenstate Thermalization Hypothesis through the use of Generalized Gibbs Ensembles \cite{DAlessio2016}, which are usually employed to characterize the behavior of integrable models. Within this framework, conservation laws are used to further restrict the statistical ensembles that determine thermal equilibrium.

\section{Conclusion}\label{SecConclusion}
We have demonstrated a general mechanism for local Hilbert space fragmentation for a large class of flat-band lattices exhibiting commutative local symmetries. These lattices obey the equitable partition theorem (EPT), which ensures the presence of CLSs and extended states with distinct parities. Upon rotating the basis, such CLSs become decoupled at the single-particle level, and in the presence of bosonic on-site interactions, we have demonstrated that this leads to the quantum fragmentation of the Hilbert space. We have shown that these lattices conserve the parity of the number of particles in all the CLSs in a single unit cell. For lattices presenting more than one set of independent local symmetries, multiple conserved quantities per unit cell can arise. Additionally, we have found the dimension of the largest sub-sector of the Hilbert space and have characterized the fragmentation as strong.

The mechanism for local Hilbert space fragmentation studied here is robust to large classes of long-range interactions, which in some cases causes the conserved quantity to change from the parity to the total number of particles in the CLSs. By studying one particular example numerically, the 1D Pyrochlore chain, we have shown that the fragmentation of the Hilbert space in these lattices leads to a nested-dome structure in the entanglement entropies. These structures can be understood through the adjacency graphs of the many-body Hamiltonian and are a consequence of the low entanglement of the CLSs. Also, thermalization is restricted to each sub-sector, which causes the entanglement entropy to be bounded by the sector-restricted Page value. 

This work leaves open the study of other classes of flat-band lattices which might present similar mechanisms of fragmentation. One could consider for instance lattices with flux, where conserved quantities were observed in the diamond chain, Creutz ladder, and dice lattice \cite{Tovmasyan2018}.

\section{Acknowledgments}
E.N. and V.A. acknowledge support through MCIN/AEI/ 10.13039/501100011033 Grant No. PID2020-118153GB-I00 and the Catalan Government (Contract No. SGR2021-00138). E.N. acknowledges financial support from MCIN/AEI/ 10.13039/501100011033 Contract No. PRE2018-085815 and from COST through Action CA16221. A.M.M. and R.G.D. acknowledge financial support from the Portuguese Institute for Nanostructures, Nanomodelling and Nanofabrication (i3N) through Projects No. UIDB/50025/2020, No. UIDP/50025/2020, and No. LA/P/0037/2020, and funding from FCT–Portuguese Foundation for Science and Technology through Project No. PTDC/FISMAC/29291/2017. A.M.M. acknowledges financial support from the FCT through the work Contract No. CDL-CTTRI147-ARH/2018 and from i3N through the work Contract No. CDL-CTTRI-46-SGRH/2022.

\begin{appendix}
\section{Level statistics and nonintegrability}\label{SecLevelStatistics}
Below we give numerical results on the level statistics of the 1D Pyrochlore chain [Fig.~\ref{FigScheme}(di)]. We consider the ordered eigenvalues $E_n$ and the associated nearest-neighbor gaps $s_n=E_{n+1}-E_n$, from which one can define the spacing ratios $r_n=\frac{\min \left(s_n, s_{n+1}\right)}{\max \left(s_n, s_{n+1}\right)}$. The level-spacing distribution of integrable systems is known to approximate a Poisson distribution, characterized by a mean value $\langle r\rangle_P=0.386$. In contrast, non-integrable systems with time-reversal symmetry tend to the probability distribution of the Gaussian orthogonal ensemble, with $\langle r\rangle_{G O E}=0.536$ \cite{Atas2013}.  In Table \ref{TableMeanSpacing}, we show the mean spacing ratio for each sub-sector $\bm{\mathcal{P}}=(\mathcal{P}_1,\cdots,\mathcal{P}_{N_c})$ of the Pyrochlore chain considered in Section \ref{SecNumerical}, with $N=4$ particles in $N_c=4$ unit cells, open boundary conditions, and one spinal site per unit cell. Most sub-sectors tend to the Wigner-Dyson value signaling non-integrability within each sub-sector. The sub-sectors belonging to $\mathcal{P}=-2$, for which the dimension of the sub-sectors is very small, tend to the Poisson regime, as those correspond to the integrable effective single-particle sub-sectors. The sectors $\mathcal{P}=-4$, not included in the table, correspond to the frozen states.

\begin{table}[t]
	\begin{center}
		\begin{tabular}{cccc}
			\hline\hline
			\,Sector $\mathcal{P}$\, & \,\,Subsector $\mathcal{P}_k$\,\, & \,\,$\langle r\rangle$\,\,& \,Standard error\,\\ \hline
			\rule{0pt}{12pt}
			$\!\!4$ & \,\,$(+,+,+,+)$\,\, & $0.520$& $0.005$\\\hline
			$2$ & \,\,$(-,+,+,+)$\,\, & $0.508$& $0.008$\\
			$2$ & \,\,$(+,-,+,+)$\,\, & $0.515$& $0.008$\\
			$2$ & \,\,$(+,+,-,+)$\,\, & $0.524$& $0.008$\\
			$2$ & \,\,$(+,+,+,-)$\,\, & $0.526$& $0.008$\\\hline
			$0$ & \,\,$(-,-,+,+)$\,\, & $0.538$& $0.014$\\
			$0$ & \,\,$(-,+,-,+)$\,\, & $0.529$& $0.014$\\
			$0$ & \,\,$(-,+,+,-)$\,\, & $0.531$& $0.014$\\
			$0$ & \,\,$(+,-,-,+)$\,\, & $0.538$& $0.013$\\
			$0$ & \,\,$(+,-,+,-)$\,\, & $0.515$& $0.014$\\
			$0$ & \,\,$(+,+,-,-)$\,\, & $0.503$& $0.013$\\\hline
			$-2$ & \,\,$(-,-,-,+)$\,\, & $0.40$& $0.03$\\
			$-2$ & \,\,$(-,-,+,-)$\,\, & $0.40$& $0.03$\\
			$-2$ & \,\,$(-,+,-,-)$\,\, & $0.36$& $0.03$\\
			$-2$ & \,\,$(+,-,-,-)$\,\, & $0.43$& $0.03$\\\hline\hline 
		\end{tabular}
	\end{center}
	\caption{Mean level spacing ratio for the sub-sectors with $\mathcal{P}=4,2,0,-2$ of the 1D Pyrochlore chain, with $N=4$ particles in $N_c=4$ unit cells, open boundary conditions, and one spinal site. The value corresponding to the Gaussian orthogonal ensemble is $\langle r\rangle_{GOE}=0.536$, while the value for a Poisson distribution is $\langle r\rangle_P=0.386$. We also indicate the standard errors of the mean. The plus and minus signs indicate the positive or negative parity of the vector $\bm{\mathcal{P}}=(\mathcal{P}_1,\cdots,\mathcal{P}_{N_c})$ in each unit cell $k$.}
	\label{TableMeanSpacing}
\end{table}

\section{Intertwinning operators}\label{SecIntertwinning}
In this subsection, we apply the argument used in \cite{Tovmasyan2018} for the Creutz ladder, diamond chain, and dice lattice with flux, to arbitrary flat-band lattices with basic commutative local symmetries of order two. The authors in \cite{Tovmasyan2018} define the so-called intertwining operators $\mathcal{U}_{j,k}$, which realize the automorphisms of the single-particle graph in the field operators. In our notation, they swap the top and bottom operators in a rung $j$ of a unit cell $k$ while leaving the other operators invariant,
\begin{equation}
	\left\lbrace\begin{aligned}
		\mathcal{U}_{j,k}\hat{t}_{j,k}\mathcal{U}_{j,k}^\dagger=\hat{b}_{j,k} \\
		\mathcal{U}_{j,k}\hat{b}_{j,k}\mathcal{U}_{j,k}^\dagger=\hat{t}_{j,k} 
	\end{aligned}\right.\, \mbox{and} \,
	\left\lbrace\begin{aligned}
		\mathcal{U}_{j,k}\hat{t}_{j'\!,k}\mathcal{U}_{j,k}^\dagger=\hat{t}_{j'\!,k} \\
		\mathcal{U}_{j,k}\hat{b}_{j'\!,k}\mathcal{U}_{j,k}^\dagger=\hat{b}_{j'\!,k} 
\end{aligned}\right.\, \mbox{for } j\neq j',
\end{equation}
while  $\mathcal{U}_{j,k}\hat{s}_{j',k'}\mathcal{U}_{j,k}^\dagger=\hat{s}_{j',k'}$ for a spinal site in any rung $j'$ and any unit cell $k'$. For each lattice, the Hamiltonian will remain invariant under a set of local permutations represented by the combined action $\mathcal{R}_k=\prod_{j=1}^{n_2/2}\mathcal{U}_{j,k}$, such that $ \mathcal{R}_k\mathcal{\hat{H}}\mathcal{R}_k^\dagger=\mathcal{\hat{H}}$. Note that, in contrast with \cite{Tovmasyan2018}, here we deal with flat-band lattices without flux. As a consequence, the intertwining operators do not cause the insertion of a $\pi$ flux and the conserved quantity does not include an additional gauge transformation. 

The effect of the intertwining operators on the compact localized states and the dispersive states we defined in Eq.~(\ref{EqCLSandDispersive}) is the following,
\begin{equation}
	\begin{aligned}
		\prod_{j_1=1}^{n_2/2}\mathcal{U}_{j_1,k}&\hat{c}_{l,k}\bigg[\!\prod_{j_2=1}^{n_2/2}\mathcal{U}_{j_2,k}\bigg]^{\!\dagger}\!\!=\\
		=&\prod_{j_1=1}^{n_2/2}\mathcal{U}_{j_1,k}\sum_{j=1}^{n_2/2} \alpha^l_{j,k}\left(\hat{t}_{j,k}-\hat{b}_{j,k}\right)\hspace{-1mm}\bigg[\prod_{j_2=1}^{n_2/2}\mathcal{U}_{j_2,k}\bigg]^{\!\dagger}\\
		=&\sum_{j=1}^{n_2/2} \alpha^l_{j,k}\left(\hat{b}_{j,k}-\hat{t}_{j,k}\right)=-\hat{c}_{l,k}.
	\end{aligned}
\end{equation}
If the intertwining operators are applied to a different unit cell, the CLS operators remain invariant $\prod_{j_1=1}^{n_2/2}\mathcal{U}_{j_1,k}\hat{c}_{l,k'}\prod_{j_2=1}^{n_2/2}\mathcal{U}_{j_2,k}^\dagger=\hat{c}_{l,k'}$, for $k\neq k'$. Due to the positive sign in the expression of the dispersive states, Eq.~(\ref{EqCLSandDispersive}), they remain invariant under such operation, $\prod_{j_1=1}^{n_2/2}\mathcal{U}_{j_1,k}\hat{d}_{l,k'}\prod_{j_2=1}^{n_2/2}\mathcal{U}_{j_2,k}^\dagger=\hat{d}_{l,k'}$ for any pair $k,k'$. For an on-site bosonic interaction, the Hamiltonian is composed of a series of terms of the form  $\hat{\rho}_{l_1,k}^{\dagger}\hat{\sigma}_{l_2,k}^{\dagger}\hat{\tau}_{l_3,k}\hat{\upsilon}_{l_4,k}$, where $\{\hat{\rho},\hat{\sigma},\hat{\tau},\hat{\upsilon}\}=\{\hat{c},\hat{d}\}$ are the annihilation operators of either a CLS or a dispersive state [see Eq.~(\ref{EqInteractionRotated})]. The terms with an odd number of CLS creation or annihilation operators change sign under the action of the intertwining operators, while the others remain invariant. As we showed in the main text, those terms always vanish, such that 
\begin{equation}
	\hat{\mathcal{P}}_k=e^{i\pi\sum_{l=1}^{n_2/2} \hat{n}_{c_{l,k}}}
\end{equation}
is a conserved quantity of the complete system.

\end{appendix}


\begin{thebibliography}{51}%
\makeatletter
\providecommand \@ifxundefined [1]{%
 \@ifx{#1\undefined}
}%
\providecommand \@ifnum [1]{%
 \ifnum #1\expandafter \@firstoftwo
 \else \expandafter \@secondoftwo
 \fi
}%
\providecommand \@ifx [1]{%
 \ifx #1\expandafter \@firstoftwo
 \else \expandafter \@secondoftwo
 \fi
}%
\providecommand \natexlab [1]{#1}%
\providecommand \enquote  [1]{``#1''}%
\providecommand \bibnamefont  [1]{#1}%
\providecommand \bibfnamefont [1]{#1}%
\providecommand \citenamefont [1]{#1}%
\providecommand \href@noop [0]{\@secondoftwo}%
\providecommand \href [0]{\begingroup \@sanitize@url \@href}%
\providecommand \@href[1]{\@@startlink{#1}\@@href}%
\providecommand \@@href[1]{\endgroup#1\@@endlink}%
\providecommand \@sanitize@url [0]{\catcode `\\12\catcode `\$12\catcode
  `\&12\catcode `\#12\catcode `\^12\catcode `\_12\catcode `\%12\relax}%
\providecommand \@@startlink[1]{}%
\providecommand \@@endlink[0]{}%
\providecommand \url  [0]{\begingroup\@sanitize@url \@url }%
\providecommand \@url [1]{\endgroup\@href {#1}{\urlprefix }}%
\providecommand \urlprefix  [0]{URL }%
\providecommand \Eprint [0]{\href }%
\providecommand \doibase [0]{https://doi.org/}%
\providecommand \selectlanguage [0]{\@gobble}%
\providecommand \bibinfo  [0]{\@secondoftwo}%
\providecommand \bibfield  [0]{\@secondoftwo}%
\providecommand \translation [1]{[#1]}%
\providecommand \BibitemOpen [0]{}%
\providecommand \bibitemStop [0]{}%
\providecommand \bibitemNoStop [0]{.\EOS\space}%
\providecommand \EOS [0]{\spacefactor3000\relax}%
\providecommand \BibitemShut  [1]{\csname bibitem#1\endcsname}%
\let\auto@bib@innerbib\@empty
\bibitem [{\citenamefont {Flach}\ \emph {et~al.}(2014)\citenamefont {Flach},
  \citenamefont {Leykam}, \citenamefont {Bodyfelt}, \citenamefont {Matthies},\
  and\ \citenamefont {Desyatnikov}}]{Flach2014a}%
  \BibitemOpen
  \bibfield  {author} {\bibinfo {author} {\bibfnamefont {S.}~\bibnamefont
  {Flach}}, \bibinfo {author} {\bibfnamefont {D.}~\bibnamefont {Leykam}},
  \bibinfo {author} {\bibfnamefont {J.~D.}\ \bibnamefont {Bodyfelt}}, \bibinfo
  {author} {\bibfnamefont {P.}~\bibnamefont {Matthies}},\ and\ \bibinfo
  {author} {\bibfnamefont {A.~S.}\ \bibnamefont {Desyatnikov}},\ }\href
  {https://doi.org/10.1209/0295-5075/105/30001} {\bibfield  {journal} {\bibinfo
   {journal} {EPL (Europhysics Letters)}\ }\textbf {\bibinfo {volume} {105}},\
  \bibinfo {pages} {30001} (\bibinfo {year} {2014})}\BibitemShut {NoStop}%
\bibitem [{\citenamefont {Maimaiti}\ \emph {et~al.}(2017)\citenamefont
  {Maimaiti}, \citenamefont {Andreanov}, \citenamefont {Park}, \citenamefont
  {Gendelman},\ and\ \citenamefont {Flach}}]{Maimati2017}%
  \BibitemOpen
  \bibfield  {author} {\bibinfo {author} {\bibfnamefont {W.}~\bibnamefont
  {Maimaiti}}, \bibinfo {author} {\bibfnamefont {A.}~\bibnamefont {Andreanov}},
  \bibinfo {author} {\bibfnamefont {H.~C.}\ \bibnamefont {Park}}, \bibinfo
  {author} {\bibfnamefont {O.}~\bibnamefont {Gendelman}},\ and\ \bibinfo
  {author} {\bibfnamefont {S.}~\bibnamefont {Flach}},\ }\href
  {https://doi.org/10.1103/PhysRevB.95.115135} {\bibfield  {journal} {\bibinfo
  {journal} {Physical Review B}\ }\textbf {\bibinfo {volume} {95}},\ \bibinfo
  {pages} {115135} (\bibinfo {year} {2017})}\BibitemShut {NoStop}%
\bibitem [{\citenamefont {Rhim}\ and\ \citenamefont {Yang}(2019)}]{Rhim2019}%
  \BibitemOpen
  \bibfield  {author} {\bibinfo {author} {\bibfnamefont {J.-W.}\ \bibnamefont
  {Rhim}}\ and\ \bibinfo {author} {\bibfnamefont {B.-J.}\ \bibnamefont
  {Yang}},\ }\href {https://doi.org/10.1103/PhysRevB.99.045107} {\bibfield
  {journal} {\bibinfo  {journal} {Physical Review B}\ }\textbf {\bibinfo
  {volume} {99}},\ \bibinfo {pages} {045107} (\bibinfo {year}
  {2019})}\BibitemShut {NoStop}%
\bibitem [{\citenamefont {Dias}\ and\ \citenamefont
  {Gouveia}(2015)}]{Dias2015}%
  \BibitemOpen
  \bibfield  {author} {\bibinfo {author} {\bibfnamefont {R.~G.}\ \bibnamefont
  {Dias}}\ and\ \bibinfo {author} {\bibfnamefont {J.~D.}\ \bibnamefont
  {Gouveia}},\ }\href {https://doi.org/10.1038/srep16852} {\bibfield  {journal}
  {\bibinfo  {journal} {Scientific Reports}\ }\textbf {\bibinfo {volume} {5}},\
  \bibinfo {pages} {16852} (\bibinfo {year} {2015})}\BibitemShut {NoStop}%
\bibitem [{\citenamefont {Nandy}\ and\ \citenamefont
  {Chakrabarti}(2015)}]{Nandy2015}%
  \BibitemOpen
  \bibfield  {author} {\bibinfo {author} {\bibfnamefont {A.}~\bibnamefont
  {Nandy}}\ and\ \bibinfo {author} {\bibfnamefont {A.}~\bibnamefont
  {Chakrabarti}},\ }\href {https://doi.org/10.1016/j.physleta.2015.09.023}
  {\bibfield  {journal} {\bibinfo  {journal} {Physics Letters, Section A:
  General, Atomic and Solid State Physics}\ }\textbf {\bibinfo {volume}
  {379}},\ \bibinfo {pages} {2876} (\bibinfo {year} {2015})}\BibitemShut
  {NoStop}%
\bibitem [{\citenamefont {Pal}\ and\ \citenamefont {Saha}(2018)}]{Pal2018}%
  \BibitemOpen
  \bibfield  {author} {\bibinfo {author} {\bibfnamefont {B.}~\bibnamefont
  {Pal}}\ and\ \bibinfo {author} {\bibfnamefont {K.}~\bibnamefont {Saha}},\
  }\href {https://doi.org/10.1103/PhysRevB.97.195101} {\bibfield  {journal}
  {\bibinfo  {journal} {Physical Review B}\ }\textbf {\bibinfo {volume} {97}},\
  \bibinfo {pages} {195101} (\bibinfo {year} {2018})}\BibitemShut {NoStop}%
\bibitem [{\citenamefont {Ramachandran}\ \emph {et~al.}(2017)\citenamefont
  {Ramachandran}, \citenamefont {Andreanov},\ and\ \citenamefont
  {Flach}}]{Ramachandran2017}%
  \BibitemOpen
  \bibfield  {author} {\bibinfo {author} {\bibfnamefont {A.}~\bibnamefont
  {Ramachandran}}, \bibinfo {author} {\bibfnamefont {A.}~\bibnamefont
  {Andreanov}},\ and\ \bibinfo {author} {\bibfnamefont {S.}~\bibnamefont
  {Flach}},\ }\href {https://doi.org/10.1103/PhysRevB.96.161104} {\bibfield
  {journal} {\bibinfo  {journal} {Physical Review B}\ }\textbf {\bibinfo
  {volume} {96}},\ \bibinfo {pages} {161104(R)} (\bibinfo {year}
  {2017})}\BibitemShut {NoStop}%
\bibitem [{\citenamefont {Morales-Inostroza}\ and\ \citenamefont
  {Vicencio}(2016)}]{MoralesInostroza2016}%
  \BibitemOpen
  \bibfield  {author} {\bibinfo {author} {\bibfnamefont {L.}~\bibnamefont
  {Morales-Inostroza}}\ and\ \bibinfo {author} {\bibfnamefont {R.~A.}\
  \bibnamefont {Vicencio}},\ }\href
  {https://doi.org/10.1103/PhysRevA.94.043831} {\bibfield  {journal} {\bibinfo
  {journal} {Physical Review A}\ }\textbf {\bibinfo {volume} {94}},\ \bibinfo
  {pages} {043831} (\bibinfo {year} {2016})}\BibitemShut {NoStop}%
\bibitem [{\citenamefont {Mielke}(1991)}]{Mielke1991}%
  \BibitemOpen
  \bibfield  {author} {\bibinfo {author} {\bibfnamefont {A.}~\bibnamefont
  {Mielke}},\ }\href {https://doi.org/10.1088/0305-4470/24/14/018} {\bibfield
  {journal} {\bibinfo  {journal} {Journal of Physics A: Mathematical and
  General}\ }\textbf {\bibinfo {volume} {24}},\ \bibinfo {pages} {3311}
  (\bibinfo {year} {1991})}\BibitemShut {NoStop}%
\bibitem [{\citenamefont {Morfonios}\ \emph {et~al.}(2021)\citenamefont
  {Morfonios}, \citenamefont {R{\"{o}}ntgen}, \citenamefont {Pyzh},\ and\
  \citenamefont {Schmelcher}}]{Morfonios2021}%
  \BibitemOpen
  \bibfield  {author} {\bibinfo {author} {\bibfnamefont {C.~V.}\ \bibnamefont
  {Morfonios}}, \bibinfo {author} {\bibfnamefont {M.}~\bibnamefont
  {R{\"{o}}ntgen}}, \bibinfo {author} {\bibfnamefont {M.}~\bibnamefont
  {Pyzh}},\ and\ \bibinfo {author} {\bibfnamefont {P.}~\bibnamefont
  {Schmelcher}},\ }\href {https://doi.org/10.1103/PhysRevB.104.035105}
  {\bibfield  {journal} {\bibinfo  {journal} {Physical Review B}\ }\textbf
  {\bibinfo {volume} {104}},\ \bibinfo {pages} {035105} (\bibinfo {year}
  {2021})}\BibitemShut {NoStop}%
\bibitem [{\citenamefont {Maimaiti}\ \emph {et~al.}(2019)\citenamefont
  {Maimaiti}, \citenamefont {Flach},\ and\ \citenamefont
  {Andreanov}}]{Maimati2021a}%
  \BibitemOpen
  \bibfield  {author} {\bibinfo {author} {\bibfnamefont {W.}~\bibnamefont
  {Maimaiti}}, \bibinfo {author} {\bibfnamefont {S.}~\bibnamefont {Flach}},\
  and\ \bibinfo {author} {\bibfnamefont {A.}~\bibnamefont {Andreanov}},\ }\href
  {https://doi.org/10.1103/PhysRevB.99.125129} {\bibfield  {journal} {\bibinfo
  {journal} {Physical Review B}\ }\textbf {\bibinfo {volume} {99}},\ \bibinfo
  {pages} {125129} (\bibinfo {year} {2019})}\BibitemShut {NoStop}%
\bibitem [{\citenamefont {Maimaiti}\ \emph {et~al.}(2021)\citenamefont
  {Maimaiti}, \citenamefont {Andreanov},\ and\ \citenamefont
  {Flach}}]{Maimati2021b}%
  \BibitemOpen
  \bibfield  {author} {\bibinfo {author} {\bibfnamefont {W.}~\bibnamefont
  {Maimaiti}}, \bibinfo {author} {\bibfnamefont {A.}~\bibnamefont
  {Andreanov}},\ and\ \bibinfo {author} {\bibfnamefont {S.}~\bibnamefont
  {Flach}},\ }\href {https://doi.org/10.1103/PhysRevB.103.165116} {\bibfield
  {journal} {\bibinfo  {journal} {Physical Review B}\ }\textbf {\bibinfo
  {volume} {103}},\ \bibinfo {pages} {165116} (\bibinfo {year}
  {2021})}\BibitemShut {NoStop}%
\bibitem [{\citenamefont {Maimaiti}\ and\ \citenamefont
  {Andreanov}(2021)}]{Maimati2021c}%
  \BibitemOpen
  \bibfield  {author} {\bibinfo {author} {\bibfnamefont {W.}~\bibnamefont
  {Maimaiti}}\ and\ \bibinfo {author} {\bibfnamefont {A.}~\bibnamefont
  {Andreanov}},\ }\href {https://doi.org/10.1103/PhysRevB.104.035115}
  {\bibfield  {journal} {\bibinfo  {journal} {Physical Review B}\ }\textbf
  {\bibinfo {volume} {104}},\ \bibinfo {pages} {035115} (\bibinfo {year}
  {2021})}\BibitemShut {NoStop}%
\bibitem [{\citenamefont {Xu}\ \emph {et~al.}(2015)\citenamefont {Xu},
  \citenamefont {Wang}, \citenamefont {Hang}, \citenamefont {Luo},
  \citenamefont {Chan},\ and\ \citenamefont {Lai}}]{Xu2015}%
  \BibitemOpen
  \bibfield  {author} {\bibinfo {author} {\bibfnamefont {C.}~\bibnamefont
  {Xu}}, \bibinfo {author} {\bibfnamefont {G.}~\bibnamefont {Wang}}, \bibinfo
  {author} {\bibfnamefont {Z.~H.}\ \bibnamefont {Hang}}, \bibinfo {author}
  {\bibfnamefont {J.}~\bibnamefont {Luo}}, \bibinfo {author} {\bibfnamefont
  {C.~T.}\ \bibnamefont {Chan}},\ and\ \bibinfo {author} {\bibfnamefont
  {Y.}~\bibnamefont {Lai}},\ }\href {https://doi.org/10.1038/srep18181}
  {\bibfield  {journal} {\bibinfo  {journal} {Scientific Reports}\ }\textbf
  {\bibinfo {volume} {5}},\ \bibinfo {pages} {18181} (\bibinfo {year}
  {2015})}\BibitemShut {NoStop}%
\bibitem [{\citenamefont {Trudeau}(1994)}]{Trudeau1994}%
  \BibitemOpen
  \bibfield  {author} {\bibinfo {author} {\bibfnamefont {R.~J.}\ \bibnamefont
  {Trudeau}},\ }\href@noop {} {\emph {\bibinfo {title} {{Introduction to Graph
  Theory}}}}\ (\bibinfo  {publisher} {Dover, Mineola, NY},\ \bibinfo {year}
  {1994})\BibitemShut {NoStop}%
\bibitem [{\citenamefont {Barrett}\ \emph {et~al.}(2017)\citenamefont
  {Barrett}, \citenamefont {Francis},\ and\ \citenamefont
  {Webb}}]{Barrett2017}%
  \BibitemOpen
  \bibfield  {author} {\bibinfo {author} {\bibfnamefont {W.}~\bibnamefont
  {Barrett}}, \bibinfo {author} {\bibfnamefont {A.}~\bibnamefont {Francis}},\
  and\ \bibinfo {author} {\bibfnamefont {B.}~\bibnamefont {Webb}},\ }\href
  {https://doi.org/10.1016/j.laa.2016.10.017} {\bibfield  {journal} {\bibinfo
  {journal} {Linear Algebra and Its Applications}\ }\textbf {\bibinfo {volume}
  {513}},\ \bibinfo {pages} {409} (\bibinfo {year} {2017})}\BibitemShut
  {NoStop}%
\bibitem [{\citenamefont {Francis}\ \emph {et~al.}(2017)\citenamefont
  {Francis}, \citenamefont {Smith}, \citenamefont {Sorensen},\ and\
  \citenamefont {Webb}}]{Francis2017}%
  \BibitemOpen
  \bibfield  {author} {\bibinfo {author} {\bibfnamefont {A.}~\bibnamefont
  {Francis}}, \bibinfo {author} {\bibfnamefont {D.}~\bibnamefont {Smith}},
  \bibinfo {author} {\bibfnamefont {D.}~\bibnamefont {Sorensen}},\ and\
  \bibinfo {author} {\bibfnamefont {B.}~\bibnamefont {Webb}},\ }\href
  {https://doi.org/10.1016/j.laa.2017.06.045} {\bibfield  {journal} {\bibinfo
  {journal} {Linear Algebra and Its Applications}\ }\textbf {\bibinfo {volume}
  {532}},\ \bibinfo {pages} {432} (\bibinfo {year} {2017})}\BibitemShut
  {NoStop}%
\bibitem [{\citenamefont {Th{\"{u}}ne}(2016)}]{Thune2016}%
  \BibitemOpen
  \bibfield  {author} {\bibinfo {author} {\bibfnamefont {M.}~\bibnamefont
  {Th{\"{u}}ne}},\ }\href {http://arxiv.org/abs/1605.05924} {\  (\bibinfo
  {year} {2016})},\ \Eprint {https://arxiv.org/abs/1605.05924}
  {arXiv:1605.05924} \BibitemShut {NoStop}%
\bibitem [{\citenamefont {R{\"{o}}ntgen}\ \emph {et~al.}(2018)\citenamefont
  {R{\"{o}}ntgen}, \citenamefont {Morfonios},\ and\ \citenamefont
  {Schmelcher}}]{Rontgen2018}%
  \BibitemOpen
  \bibfield  {author} {\bibinfo {author} {\bibfnamefont {M.}~\bibnamefont
  {R{\"{o}}ntgen}}, \bibinfo {author} {\bibfnamefont {C.~V.}\ \bibnamefont
  {Morfonios}},\ and\ \bibinfo {author} {\bibfnamefont {P.}~\bibnamefont
  {Schmelcher}},\ }\href {https://doi.org/10.1103/PhysRevB.97.035161}
  {\bibfield  {journal} {\bibinfo  {journal} {Physical Review B}\ }\textbf
  {\bibinfo {volume} {97}},\ \bibinfo {pages} {035161} (\bibinfo {year}
  {2018})}\BibitemShut {NoStop}%
\bibitem [{\citenamefont {Nicolau}\ \emph
  {et~al.}(2023{\natexlab{a}})\citenamefont {Nicolau}, \citenamefont {Marques},
  \citenamefont {Mompart}, \citenamefont {Ahufinger},\ and\ \citenamefont
  {Dias}}]{Nicolau2022a}%
  \BibitemOpen
  \bibfield  {author} {\bibinfo {author} {\bibfnamefont {E.}~\bibnamefont
  {Nicolau}}, \bibinfo {author} {\bibfnamefont {A.~M.}\ \bibnamefont
  {Marques}}, \bibinfo {author} {\bibfnamefont {J.}~\bibnamefont {Mompart}},
  \bibinfo {author} {\bibfnamefont {V.}~\bibnamefont {Ahufinger}},\ and\
  \bibinfo {author} {\bibfnamefont {R.~G.}\ \bibnamefont {Dias}},\ }\href
  {https://doi.org/10.1103/PhysRevB.107.094312} {\bibfield  {journal} {\bibinfo
   {journal} {Physical Review B}\ }\textbf {\bibinfo {volume} {107}},\ \bibinfo
  {pages} {094312} (\bibinfo {year} {2023}{\natexlab{a}})}\BibitemShut
  {NoStop}%
\bibitem [{\citenamefont {Bu{\v{c}}a}(2022)}]{Buca2022}%
  \BibitemOpen
  \bibfield  {author} {\bibinfo {author} {\bibfnamefont {B.}~\bibnamefont
  {Bu{\v{c}}a}},\ }\href {https://doi.org/10.1103/PhysRevLett.128.100601}
  {\bibfield  {journal} {\bibinfo  {journal} {Physical Review Letters}\
  }\textbf {\bibinfo {volume} {128}},\ \bibinfo {pages} {100601} (\bibinfo
  {year} {2022})}\BibitemShut {NoStop}%
\bibitem [{\citenamefont {Mukherjee}\ \emph
  {et~al.}(2021{\natexlab{a}})\citenamefont {Mukherjee}, \citenamefont
  {Banerjee}, \citenamefont {Sengupta},\ and\ \citenamefont
  {Sen}}]{Mukherjee2021}%
  \BibitemOpen
  \bibfield  {author} {\bibinfo {author} {\bibfnamefont {B.}~\bibnamefont
  {Mukherjee}}, \bibinfo {author} {\bibfnamefont {D.}~\bibnamefont {Banerjee}},
  \bibinfo {author} {\bibfnamefont {K.}~\bibnamefont {Sengupta}},\ and\
  \bibinfo {author} {\bibfnamefont {A.}~\bibnamefont {Sen}},\ }\href
  {https://doi.org/10.1103/PhysRevB.104.155117} {\bibfield  {journal} {\bibinfo
   {journal} {Physical Review B}\ }\textbf {\bibinfo {volume} {104}},\ \bibinfo
  {pages} {155117} (\bibinfo {year} {2021}{\natexlab{a}})}\BibitemShut
  {NoStop}%
\bibitem [{\citenamefont {Danieli}\ \emph {et~al.}(2020)\citenamefont
  {Danieli}, \citenamefont {Andreanov},\ and\ \citenamefont
  {Flach}}]{Danieli2020}%
  \BibitemOpen
  \bibfield  {author} {\bibinfo {author} {\bibfnamefont {C.}~\bibnamefont
  {Danieli}}, \bibinfo {author} {\bibfnamefont {A.}~\bibnamefont {Andreanov}},\
  and\ \bibinfo {author} {\bibfnamefont {S.}~\bibnamefont {Flach}},\ }\href
  {https://doi.org/10.1103/PhysRevB.102.041116} {\bibfield  {journal} {\bibinfo
   {journal} {Physical Review B}\ }\textbf {\bibinfo {volume} {102}},\ \bibinfo
  {pages} {041116(R)} (\bibinfo {year} {2020})}\BibitemShut {NoStop}%
\bibitem [{\citenamefont {Hahn}\ \emph {et~al.}(2021)\citenamefont {Hahn},
  \citenamefont {McClarty},\ and\ \citenamefont {Luitz}}]{Hahn2021}%
  \BibitemOpen
  \bibfield  {author} {\bibinfo {author} {\bibfnamefont {D.}~\bibnamefont
  {Hahn}}, \bibinfo {author} {\bibfnamefont {P.~A.}\ \bibnamefont {McClarty}},\
  and\ \bibinfo {author} {\bibfnamefont {D.~J.}\ \bibnamefont {Luitz}},\ }\href
  {https://doi.org/10.21468/SciPostPhys.11.4.074} {\bibfield  {journal}
  {\bibinfo  {journal} {SciPost Physics}\ }\textbf {\bibinfo {volume} {11}},\
  \bibinfo {pages} {074} (\bibinfo {year} {2021})}\BibitemShut {NoStop}%
\bibitem [{\citenamefont {Chertkov}\ and\ \citenamefont
  {Clark}(2021)}]{Chertkov2021}%
  \BibitemOpen
  \bibfield  {author} {\bibinfo {author} {\bibfnamefont {E.}~\bibnamefont
  {Chertkov}}\ and\ \bibinfo {author} {\bibfnamefont {B.~K.}\ \bibnamefont
  {Clark}},\ }\href {https://doi.org/10.1103/PhysRevB.104.104410} {\bibfield
  {journal} {\bibinfo  {journal} {Physical Review B}\ }\textbf {\bibinfo
  {volume} {104}},\ \bibinfo {pages} {104410} (\bibinfo {year}
  {2021})}\BibitemShut {NoStop}%
\bibitem [{\citenamefont {Smith}\ \emph {et~al.}(2017)\citenamefont {Smith},
  \citenamefont {Knolle}, \citenamefont {Kovrizhin},\ and\ \citenamefont
  {Moessner}}]{Smith2017}%
  \BibitemOpen
  \bibfield  {author} {\bibinfo {author} {\bibfnamefont {A.}~\bibnamefont
  {Smith}}, \bibinfo {author} {\bibfnamefont {J.}~\bibnamefont {Knolle}},
  \bibinfo {author} {\bibfnamefont {D.~L.}\ \bibnamefont {Kovrizhin}},\ and\
  \bibinfo {author} {\bibfnamefont {R.}~\bibnamefont {Moessner}},\ }\href
  {https://doi.org/10.1103/PhysRevLett.118.266601} {\bibfield  {journal}
  {\bibinfo  {journal} {Physical Review Letters}\ }\textbf {\bibinfo {volume}
  {118}},\ \bibinfo {pages} {266601} (\bibinfo {year} {2017})}\BibitemShut
  {NoStop}%
\bibitem [{\citenamefont {Smith}\ \emph {et~al.}(2018)\citenamefont {Smith},
  \citenamefont {Knolle}, \citenamefont {Moessner},\ and\ \citenamefont
  {Kovrizhin}}]{Smith2018}%
  \BibitemOpen
  \bibfield  {author} {\bibinfo {author} {\bibfnamefont {A.}~\bibnamefont
  {Smith}}, \bibinfo {author} {\bibfnamefont {J.}~\bibnamefont {Knolle}},
  \bibinfo {author} {\bibfnamefont {R.}~\bibnamefont {Moessner}},\ and\
  \bibinfo {author} {\bibfnamefont {D.~L.}\ \bibnamefont {Kovrizhin}},\ }\href
  {https://doi.org/10.1103/PhysRevB.97.245137} {\bibfield  {journal} {\bibinfo
  {journal} {Physical Review B}\ }\textbf {\bibinfo {volume} {97}},\ \bibinfo
  {pages} {245137} (\bibinfo {year} {2018})}\BibitemShut {NoStop}%
\bibitem [{\citenamefont {Russomanno}\ \emph {et~al.}(2020)\citenamefont
  {Russomanno}, \citenamefont {Notarnicola}, \citenamefont {Surace},
  \citenamefont {Fazio}, \citenamefont {Dalmonte},\ and\ \citenamefont
  {Heyl}}]{Russomanno2020b}%
  \BibitemOpen
  \bibfield  {author} {\bibinfo {author} {\bibfnamefont {A.}~\bibnamefont
  {Russomanno}}, \bibinfo {author} {\bibfnamefont {S.}~\bibnamefont
  {Notarnicola}}, \bibinfo {author} {\bibfnamefont {F.~M.}\ \bibnamefont
  {Surace}}, \bibinfo {author} {\bibfnamefont {R.}~\bibnamefont {Fazio}},
  \bibinfo {author} {\bibfnamefont {M.}~\bibnamefont {Dalmonte}},\ and\
  \bibinfo {author} {\bibfnamefont {M.}~\bibnamefont {Heyl}},\ }\href
  {https://doi.org/10.1103/PhysRevResearch.2.012003} {\bibfield  {journal}
  {\bibinfo  {journal} {Physical Review Research}\ }\textbf {\bibinfo {volume}
  {2}},\ \bibinfo {pages} {012003(R)} (\bibinfo {year} {2020})}\BibitemShut
  {NoStop}%
\bibitem [{\citenamefont {Halimeh}\ \emph {et~al.}(2022)\citenamefont
  {Halimeh}, \citenamefont {Hauke}, \citenamefont {Knolle},\ and\ \citenamefont
  {Grusdt}}]{Halimeh2022}%
  \BibitemOpen
  \bibfield  {author} {\bibinfo {author} {\bibfnamefont {J.~C.}\ \bibnamefont
  {Halimeh}}, \bibinfo {author} {\bibfnamefont {P.}~\bibnamefont {Hauke}},
  \bibinfo {author} {\bibfnamefont {J.}~\bibnamefont {Knolle}},\ and\ \bibinfo
  {author} {\bibfnamefont {F.}~\bibnamefont {Grusdt}},\ }\href
  {http://arxiv.org/abs/2206.11273} {\  (\bibinfo {year} {2022})},\ \Eprint
  {https://arxiv.org/abs/2206.11273} {arXiv:2206.11273} \BibitemShut {NoStop}%
\bibitem [{\citenamefont {Borla}\ \emph {et~al.}(2020)\citenamefont {Borla},
  \citenamefont {Verresen}, \citenamefont {Grusdt},\ and\ \citenamefont
  {Moroz}}]{Borla2020}%
  \BibitemOpen
  \bibfield  {author} {\bibinfo {author} {\bibfnamefont {U.}~\bibnamefont
  {Borla}}, \bibinfo {author} {\bibfnamefont {R.}~\bibnamefont {Verresen}},
  \bibinfo {author} {\bibfnamefont {F.}~\bibnamefont {Grusdt}},\ and\ \bibinfo
  {author} {\bibfnamefont {S.}~\bibnamefont {Moroz}},\ }\href
  {https://doi.org/10.1103/PhysRevLett.124.120503} {\bibfield  {journal}
  {\bibinfo  {journal} {Physical Review Letters}\ }\textbf {\bibinfo {volume}
  {124}},\ \bibinfo {pages} {120503} (\bibinfo {year} {2020})}\BibitemShut
  {NoStop}%
\bibitem [{\citenamefont {Bu{\v{c}}a}(2023)}]{Buca2023}%
  \BibitemOpen
  \bibfield  {author} {\bibinfo {author} {\bibfnamefont {B.}~\bibnamefont
  {Bu{\v{c}}a}},\ }\href {http://arxiv.org/abs/2301.07091} {\  (\bibinfo {year}
  {2023})},\ \Eprint {https://arxiv.org/abs/2301.07091} {arXiv:2301.07091}
  \BibitemShut {NoStop}%
\bibitem [{\citenamefont {Marques}\ \emph {et~al.}(2021)\citenamefont
  {Marques}, \citenamefont {Madail},\ and\ \citenamefont {Dias}}]{Marques2021}%
  \BibitemOpen
  \bibfield  {author} {\bibinfo {author} {\bibfnamefont {A.~M.}\ \bibnamefont
  {Marques}}, \bibinfo {author} {\bibfnamefont {L.}~\bibnamefont {Madail}},\
  and\ \bibinfo {author} {\bibfnamefont {R.~G.}\ \bibnamefont {Dias}},\ }\href
  {https://doi.org/10.1103/PhysRevB.103.235425} {\bibfield  {journal} {\bibinfo
   {journal} {Physical Review B}\ }\textbf {\bibinfo {volume} {103}},\ \bibinfo
  {pages} {235425} (\bibinfo {year} {2021})}\BibitemShut {NoStop}%
\bibitem [{\citenamefont {McClarty}\ \emph {et~al.}(2020)\citenamefont
  {McClarty}, \citenamefont {Haque}, \citenamefont {Sen},\ and\ \citenamefont
  {Richter}}]{McClarty2020}%
  \BibitemOpen
  \bibfield  {author} {\bibinfo {author} {\bibfnamefont {P.~A.}\ \bibnamefont
  {McClarty}}, \bibinfo {author} {\bibfnamefont {M.}~\bibnamefont {Haque}},
  \bibinfo {author} {\bibfnamefont {A.}~\bibnamefont {Sen}},\ and\ \bibinfo
  {author} {\bibfnamefont {J.}~\bibnamefont {Richter}},\ }\href
  {https://doi.org/10.1103/PhysRevB.102.224303} {\bibfield  {journal} {\bibinfo
   {journal} {Physical Review B}\ }\textbf {\bibinfo {volume} {102}},\ \bibinfo
  {pages} {224303} (\bibinfo {year} {2020})}\BibitemShut {NoStop}%
\bibitem [{\citenamefont {Lopes}\ and\ \citenamefont {Dias}(2011)}]{Lopes2011}%
  \BibitemOpen
  \bibfield  {author} {\bibinfo {author} {\bibfnamefont {A.~A.}\ \bibnamefont
  {Lopes}}\ and\ \bibinfo {author} {\bibfnamefont {R.~G.}\ \bibnamefont
  {Dias}},\ }\href {https://doi.org/10.1103/PhysRevB.84.085124} {\bibfield
  {journal} {\bibinfo  {journal} {Physical Review B}\ }\textbf {\bibinfo
  {volume} {84}},\ \bibinfo {pages} {085124} (\bibinfo {year}
  {2011})}\BibitemShut {NoStop}%
\bibitem [{\citenamefont {Pelegr{\'{i}}}\ \emph
  {et~al.}(2019{\natexlab{a}})\citenamefont {Pelegr{\'{i}}}, \citenamefont
  {Marques}, \citenamefont {Dias}, \citenamefont {Daley}, \citenamefont
  {Ahufinger},\ and\ \citenamefont {Mompart}}]{Pelegri2019}%
  \BibitemOpen
  \bibfield  {author} {\bibinfo {author} {\bibfnamefont {G.}~\bibnamefont
  {Pelegr{\'{i}}}}, \bibinfo {author} {\bibfnamefont {A.~M.}\ \bibnamefont
  {Marques}}, \bibinfo {author} {\bibfnamefont {R.~G.}\ \bibnamefont {Dias}},
  \bibinfo {author} {\bibfnamefont {A.~J.}\ \bibnamefont {Daley}}, \bibinfo
  {author} {\bibfnamefont {V.}~\bibnamefont {Ahufinger}},\ and\ \bibinfo
  {author} {\bibfnamefont {J.}~\bibnamefont {Mompart}},\ }\href
  {https://doi.org/10.1103/PhysRevA.99.023612} {\bibfield  {journal} {\bibinfo
  {journal} {Physical Review A}\ }\textbf {\bibinfo {volume} {99}},\ \bibinfo
  {pages} {023612} (\bibinfo {year} {2019}{\natexlab{a}})}\BibitemShut
  {NoStop}%
\bibitem [{\citenamefont {Pelegr{\'{i}}}\ \emph
  {et~al.}(2019{\natexlab{b}})\citenamefont {Pelegr{\'{i}}}, \citenamefont
  {Marques}, \citenamefont {Dias}, \citenamefont {Daley}, \citenamefont
  {Mompart},\ and\ \citenamefont {Ahufinger}}]{Pelegri2019d}%
  \BibitemOpen
  \bibfield  {author} {\bibinfo {author} {\bibfnamefont {G.}~\bibnamefont
  {Pelegr{\'{i}}}}, \bibinfo {author} {\bibfnamefont {A.~M.}\ \bibnamefont
  {Marques}}, \bibinfo {author} {\bibfnamefont {R.~G.}\ \bibnamefont {Dias}},
  \bibinfo {author} {\bibfnamefont {A.~J.}\ \bibnamefont {Daley}}, \bibinfo
  {author} {\bibfnamefont {J.}~\bibnamefont {Mompart}},\ and\ \bibinfo {author}
  {\bibfnamefont {V.}~\bibnamefont {Ahufinger}},\ }\href
  {https://doi.org/10.1103/PhysRevA.99.023613} {\bibfield  {journal} {\bibinfo
  {journal} {Physical Review A}\ }\textbf {\bibinfo {volume} {99}},\ \bibinfo
  {pages} {023613} (\bibinfo {year} {2019}{\natexlab{b}})}\BibitemShut
  {NoStop}%
\bibitem [{\citenamefont {Mukherjee}\ \emph
  {et~al.}(2021{\natexlab{b}})\citenamefont {Mukherjee}, \citenamefont {Nandy},
  \citenamefont {Sil},\ and\ \citenamefont {Chakrabarti}}]{Mukherjee2020}%
  \BibitemOpen
  \bibfield  {author} {\bibinfo {author} {\bibfnamefont {A.}~\bibnamefont
  {Mukherjee}}, \bibinfo {author} {\bibfnamefont {A.}~\bibnamefont {Nandy}},
  \bibinfo {author} {\bibfnamefont {S.}~\bibnamefont {Sil}},\ and\ \bibinfo
  {author} {\bibfnamefont {A.}~\bibnamefont {Chakrabarti}},\ }\href
  {https://doi.org/10.1088/1361-648X/abbc9a} {\bibfield  {journal} {\bibinfo
  {journal} {Journal of Physics: Condensed Matter}\ }\textbf {\bibinfo {volume}
  {33}},\ \bibinfo {pages} {035502} (\bibinfo {year}
  {2021}{\natexlab{b}})}\BibitemShut {NoStop}%
\bibitem [{\citenamefont {Nicolau}\ \emph
  {et~al.}(2023{\natexlab{b}})\citenamefont {Nicolau}, \citenamefont {Marques},
  \citenamefont {Dias}, \citenamefont {Mompart},\ and\ \citenamefont
  {Ahufinger}}]{Nicolau2022}%
  \BibitemOpen
  \bibfield  {author} {\bibinfo {author} {\bibfnamefont {E.}~\bibnamefont
  {Nicolau}}, \bibinfo {author} {\bibfnamefont {A.~M.}\ \bibnamefont
  {Marques}}, \bibinfo {author} {\bibfnamefont {R.~G.}\ \bibnamefont {Dias}},
  \bibinfo {author} {\bibfnamefont {J.}~\bibnamefont {Mompart}},\ and\ \bibinfo
  {author} {\bibfnamefont {V.}~\bibnamefont {Ahufinger}},\ }\href
  {https://doi.org/10.1103/PhysRevA.107.023305} {\bibfield  {journal} {\bibinfo
   {journal} {Physical Review A}\ }\textbf {\bibinfo {volume} {107}},\ \bibinfo
  {pages} {023305} (\bibinfo {year} {2023}{\natexlab{b}})}\BibitemShut
  {NoStop}%
\bibitem [{\citenamefont {Tovmasyan}\ \emph {et~al.}(2013)\citenamefont
  {Tovmasyan}, \citenamefont {van Nieuwenburg},\ and\ \citenamefont
  {Huber}}]{Tovmasyan2013}%
  \BibitemOpen
  \bibfield  {author} {\bibinfo {author} {\bibfnamefont {M.}~\bibnamefont
  {Tovmasyan}}, \bibinfo {author} {\bibfnamefont {E.~P.~L.}\ \bibnamefont {van
  Nieuwenburg}},\ and\ \bibinfo {author} {\bibfnamefont {S.~D.}\ \bibnamefont
  {Huber}},\ }\href {https://doi.org/10.1103/PhysRevB.88.220510} {\bibfield
  {journal} {\bibinfo  {journal} {Physical Review B}\ }\textbf {\bibinfo
  {volume} {88}},\ \bibinfo {pages} {220510(R)} (\bibinfo {year}
  {2013})}\BibitemShut {NoStop}%
\bibitem [{\citenamefont {Pelegr{\'{i}}}\ \emph {et~al.}(2020)\citenamefont
  {Pelegr{\'{i}}}, \citenamefont {Marques}, \citenamefont {Ahufinger},
  \citenamefont {Mompart},\ and\ \citenamefont {Dias}}]{Pelegri2020}%
  \BibitemOpen
  \bibfield  {author} {\bibinfo {author} {\bibfnamefont {G.}~\bibnamefont
  {Pelegr{\'{i}}}}, \bibinfo {author} {\bibfnamefont {A.~M.}\ \bibnamefont
  {Marques}}, \bibinfo {author} {\bibfnamefont {V.}~\bibnamefont {Ahufinger}},
  \bibinfo {author} {\bibfnamefont {J.}~\bibnamefont {Mompart}},\ and\ \bibinfo
  {author} {\bibfnamefont {R.~G.}\ \bibnamefont {Dias}},\ }\href
  {https://doi.org/10.1103/PhysRevResearch.2.033267} {\bibfield  {journal}
  {\bibinfo  {journal} {Physical Review Research}\ }\textbf {\bibinfo {volume}
  {2}},\ \bibinfo {pages} {033267} (\bibinfo {year} {2020})}\BibitemShut
  {NoStop}%
\bibitem [{\citenamefont {Kuno}\ \emph {et~al.}(2020)\citenamefont {Kuno},
  \citenamefont {Mizoguchi},\ and\ \citenamefont {Hatsugai}}]{Kuno2020a}%
  \BibitemOpen
  \bibfield  {author} {\bibinfo {author} {\bibfnamefont {Y.}~\bibnamefont
  {Kuno}}, \bibinfo {author} {\bibfnamefont {T.}~\bibnamefont {Mizoguchi}},\
  and\ \bibinfo {author} {\bibfnamefont {Y.}~\bibnamefont {Hatsugai}},\ }\href
  {https://doi.org/10.1103/PhysRevA.102.063325} {\bibfield  {journal} {\bibinfo
   {journal} {Physical Review A}\ }\textbf {\bibinfo {volume} {102}},\ \bibinfo
  {pages} {063325} (\bibinfo {year} {2020})}\BibitemShut {NoStop}%
\bibitem [{\citenamefont {Moudgalya}\ and\ \citenamefont
  {Motrunich}(2022)}]{Moudgalya2021b}%
  \BibitemOpen
  \bibfield  {author} {\bibinfo {author} {\bibfnamefont {S.}~\bibnamefont
  {Moudgalya}}\ and\ \bibinfo {author} {\bibfnamefont {O.~I.}\ \bibnamefont
  {Motrunich}},\ }\href {https://doi.org/10.1103/PhysRevX.12.011050} {\bibfield
   {journal} {\bibinfo  {journal} {Physical Review X}\ }\textbf {\bibinfo
  {volume} {12}},\ \bibinfo {pages} {011050} (\bibinfo {year}
  {2022})}\BibitemShut {NoStop}%
\bibitem [{\citenamefont {Brighi}\ \emph {et~al.}(2022)\citenamefont {Brighi},
  \citenamefont {Ljubotina},\ and\ \citenamefont {Serbyn}}]{Brighi2022b}%
  \BibitemOpen
  \bibfield  {author} {\bibinfo {author} {\bibfnamefont {P.}~\bibnamefont
  {Brighi}}, \bibinfo {author} {\bibfnamefont {M.}~\bibnamefont {Ljubotina}},\
  and\ \bibinfo {author} {\bibfnamefont {M.}~\bibnamefont {Serbyn}},\ }\href
  {http://arxiv.org/abs/2210.15607} {\  (\bibinfo {year} {2022})},\ \Eprint
  {https://arxiv.org/abs/2210.15607} {arXiv:2210.15607} \BibitemShut {NoStop}%
\bibitem [{\citenamefont {Borsi}\ \emph {et~al.}(2023)\citenamefont {Borsi},
  \citenamefont {Pristy{\'{a}}k},\ and\ \citenamefont {Pozsgay}}]{Borsi2023}%
  \BibitemOpen
  \bibfield  {author} {\bibinfo {author} {\bibfnamefont {M.}~\bibnamefont
  {Borsi}}, \bibinfo {author} {\bibfnamefont {L.}~\bibnamefont
  {Pristy{\'{a}}k}},\ and\ \bibinfo {author} {\bibfnamefont {B.}~\bibnamefont
  {Pozsgay}},\ }\href {http://arxiv.org/abs/2302.07219} {\  (\bibinfo {year}
  {2023})},\ \Eprint {https://arxiv.org/abs/2302.07219} {arXiv:2302.07219}
  \BibitemShut {NoStop}%
\bibitem [{\citenamefont {Moudgalya}\ \emph {et~al.}(2022)\citenamefont
  {Moudgalya}, \citenamefont {Bernevig},\ and\ \citenamefont
  {Regnault}}]{Moudgalya2021a}%
  \BibitemOpen
  \bibfield  {author} {\bibinfo {author} {\bibfnamefont {S.}~\bibnamefont
  {Moudgalya}}, \bibinfo {author} {\bibfnamefont {B.~A.}\ \bibnamefont
  {Bernevig}},\ and\ \bibinfo {author} {\bibfnamefont {N.}~\bibnamefont
  {Regnault}},\ }\href {https://doi.org/10.1088/1361-6633/ac73a0} {\bibfield
  {journal} {\bibinfo  {journal} {Reports on Progress in Physics}\ }\textbf
  {\bibinfo {volume} {85}},\ \bibinfo {pages} {086501} (\bibinfo {year}
  {2022})}\BibitemShut {NoStop}%
\bibitem [{\citenamefont {Sala}\ \emph {et~al.}(2020)\citenamefont {Sala},
  \citenamefont {Rakovszky}, \citenamefont {Verresen}, \citenamefont {Knap},\
  and\ \citenamefont {Pollmann}}]{Sala2020}%
  \BibitemOpen
  \bibfield  {author} {\bibinfo {author} {\bibfnamefont {P.}~\bibnamefont
  {Sala}}, \bibinfo {author} {\bibfnamefont {T.}~\bibnamefont {Rakovszky}},
  \bibinfo {author} {\bibfnamefont {R.}~\bibnamefont {Verresen}}, \bibinfo
  {author} {\bibfnamefont {M.}~\bibnamefont {Knap}},\ and\ \bibinfo {author}
  {\bibfnamefont {F.}~\bibnamefont {Pollmann}},\ }\href
  {https://doi.org/10.1103/PhysRevX.10.011047} {\bibfield  {journal} {\bibinfo
  {journal} {Physical Review X}\ }\textbf {\bibinfo {volume} {10}},\ \bibinfo
  {pages} {011047} (\bibinfo {year} {2020})}\BibitemShut {NoStop}%
\bibitem [{\citenamefont {Tovmasyan}\ \emph {et~al.}(2018)\citenamefont
  {Tovmasyan}, \citenamefont {Peotta}, \citenamefont {Liang}, \citenamefont
  {T{\"{o}}rm{\"{a}}},\ and\ \citenamefont {Huber}}]{Tovmasyan2018}%
  \BibitemOpen
  \bibfield  {author} {\bibinfo {author} {\bibfnamefont {M.}~\bibnamefont
  {Tovmasyan}}, \bibinfo {author} {\bibfnamefont {S.}~\bibnamefont {Peotta}},
  \bibinfo {author} {\bibfnamefont {L.}~\bibnamefont {Liang}}, \bibinfo
  {author} {\bibfnamefont {P.}~\bibnamefont {T{\"{o}}rm{\"{a}}}},\ and\
  \bibinfo {author} {\bibfnamefont {S.~D.}\ \bibnamefont {Huber}},\ }\href
  {https://doi.org/10.1103/PhysRevB.98.134513} {\bibfield  {journal} {\bibinfo
  {journal} {Physical Review B}\ }\textbf {\bibinfo {volume} {98}},\ \bibinfo
  {pages} {134513} (\bibinfo {year} {2018})}\BibitemShut {NoStop}%
\bibitem [{\citenamefont {Danieli}\ \emph {et~al.}(2021)\citenamefont
  {Danieli}, \citenamefont {Andreanov}, \citenamefont {Mithun},\ and\
  \citenamefont {Flach}}]{Danieli2021}%
  \BibitemOpen
  \bibfield  {author} {\bibinfo {author} {\bibfnamefont {C.}~\bibnamefont
  {Danieli}}, \bibinfo {author} {\bibfnamefont {A.}~\bibnamefont {Andreanov}},
  \bibinfo {author} {\bibfnamefont {T.}~\bibnamefont {Mithun}},\ and\ \bibinfo
  {author} {\bibfnamefont {S.}~\bibnamefont {Flach}},\ }\href
  {https://doi.org/10.1103/PhysRevB.104.085132} {\bibfield  {journal} {\bibinfo
   {journal} {Physical Review B}\ }\textbf {\bibinfo {volume} {104}},\ \bibinfo
  {pages} {085132} (\bibinfo {year} {2021})}\BibitemShut {NoStop}%
\bibitem [{\citenamefont {Page}(1993)}]{Page1993}%
  \BibitemOpen
  \bibfield  {author} {\bibinfo {author} {\bibfnamefont {D.~N.}\ \bibnamefont
  {Page}},\ }\href {https://doi.org/10.1103/PhysRevLett.71.1291} {\bibfield
  {journal} {\bibinfo  {journal} {Physical Review Letters}\ }\textbf {\bibinfo
  {volume} {71}},\ \bibinfo {pages} {1291} (\bibinfo {year}
  {1993})}\BibitemShut {NoStop}%
\bibitem [{\citenamefont {D'Alessio}\ \emph {et~al.}(2016)\citenamefont
  {D'Alessio}, \citenamefont {Kafri}, \citenamefont {Polkovnikov},\ and\
  \citenamefont {Rigol}}]{DAlessio2016}%
  \BibitemOpen
  \bibfield  {author} {\bibinfo {author} {\bibfnamefont {L.}~\bibnamefont
  {D'Alessio}}, \bibinfo {author} {\bibfnamefont {Y.}~\bibnamefont {Kafri}},
  \bibinfo {author} {\bibfnamefont {A.}~\bibnamefont {Polkovnikov}},\ and\
  \bibinfo {author} {\bibfnamefont {M.}~\bibnamefont {Rigol}},\ }\href
  {https://doi.org/10.1080/00018732.2016.1198134} {\bibfield  {journal}
  {\bibinfo  {journal} {Advances in Physics}\ }\textbf {\bibinfo {volume}
  {65}},\ \bibinfo {pages} {239} (\bibinfo {year} {2016})}\BibitemShut
  {NoStop}%
\bibitem [{\citenamefont {Atas}\ \emph {et~al.}(2013)\citenamefont {Atas},
  \citenamefont {Bogomolny}, \citenamefont {Giraud},\ and\ \citenamefont
  {Roux}}]{Atas2013}%
  \BibitemOpen
  \bibfield  {author} {\bibinfo {author} {\bibfnamefont {Y.~Y.}\ \bibnamefont
  {Atas}}, \bibinfo {author} {\bibfnamefont {E.}~\bibnamefont {Bogomolny}},
  \bibinfo {author} {\bibfnamefont {O.}~\bibnamefont {Giraud}},\ and\ \bibinfo
  {author} {\bibfnamefont {G.}~\bibnamefont {Roux}},\ }\href
  {https://doi.org/10.1103/PhysRevLett.110.084101} {\bibfield  {journal}
  {\bibinfo  {journal} {Physical Review Letters}\ }\textbf {\bibinfo {volume}
  {110}},\ \bibinfo {pages} {084101} (\bibinfo {year} {2013})}\BibitemShut
  {NoStop}%
\end{thebibliography}
\end{document}